\documentclass[11pt]{article}
\usepackage{comment,soul}
\usepackage[hidelinks]{hyperref}
\usepackage{enumitem}
\usepackage{amsmath,amssymb,epsf,cite,graphicx,subfigure,faktor,slashed}
\usepackage{amsmath,braket,tikzsymbols}
\usepackage{mathrsfs}
\usepackage[bbgreekl]{mathbbol}
\usepackage{comment}
\usepackage{dsfont}

\numberwithin{equation}{section}

\definecolor{airforceblue}{rgb}{0.36, 0.54, 0.66}
\newcommand{\beq}{\begin{equation}}
\newcommand{\eeq}{\end{equation}}

\usepackage{todonotes}

\usepackage{simplewick}

\begin{document}
\baselineskip=15.5pt
\pagestyle{plain}
\setcounter{page}{1}

\begin{center}

{\LARGE \bf Soft gravitons in three dimensions}
\vskip 1cm

\textbf{Jordan Cotler$^{1,a}$, Kristan Jensen$^{2,b}$, Stefan Prohazka$^{3,c}$, \\ Max Riegler$^{4,d}$, and Jakob Salzer$^{5,e}$}

\vspace{0.5cm}

{\it ${}^1$ Department of Physics, Harvard University, Cambridge, MA 02138, USA \vspace{.3cm}}

{\it${}^2$Department of Physics and Astronomy, University of Victoria, Victoria, BC V8W 3P6, Canada\vspace{.3cm}}

{\it ${}^3$University of Vienna, Faculty of Physics, Mathematical Physics, \\ Boltzmanngasse 5, 1090, Vienna, Austria\vspace{.3cm}}

{\it ${}^4$Quantum Technology Laboratories GmbH, \\ Clemens-Holzmeister-Straße 6/6, 1100 Vienna, Austria\vspace{.3cm}}

{\it ${}^5$Universit\'e Libre de Bruxelles and International Solvay Institutes, \\ ULB-Campus Plaine CP231, B-1050 Brussels, Belgium}

\vspace{0.3cm}

{\tt ${}^a$jcotler@fas.harvard.edu, \small ${}^b$kristanj@uvic.ca, ${}^c$stefan.prohazka@univie.ac.at, ${}^d$rieglerm@hep.itp.tuwien.ac.at, ${}^e$jakob.salzer@ulb.be\\}

\medskip

\end{center}

\vskip1cm

\begin{center}
{\bf Abstract}
\end{center}
\hspace{.3cm} 
We consider quantum gravity with zero cosmological constant in three dimensions.  First, we show that pure quantum gravity can be written as a magnetic Carrollian theory living on null infinity, described by Schwarzian-like degrees of freedom.  Next, we couple quantum gravity to massless matter. Transition amplitudes exhibit several features that resemble soft graviton physics in four dimensions, despite the absence of a propagating graviton. As in four dimensions, we find three equivalent results: a soft graviton theorem, an infinite-dimensional BMS asymptotic symmetry, and a gravitational memory effect. We also resolve some extant puzzles concerning the partition function and Hilbert space of pure 3d gravity with zero cosmological constant.

\newpage

\tableofcontents

\section{Introduction}
Consider four-dimensional quantum field theory coupled to perturbative quantum gravity with vanishing cosmological constant. Weinberg's soft graviton theorem~\cite{Weinberg:1965nx} describes a universal contribution to the $S$-matrix coming from very long wavelength gravitons. Despite its age, the soft graviton theorem (and its cousins) continue to be a focal point of ongoing research. In modern terms the soft graviton theorem is equivalent to two other features of flat space perturbative quantum gravity, namely the gravitational memory effect~\cite{Braginsky:1987kwo} and the infinite-dimensional BMS asymptotic symmetry of flat space gravity~\cite{Bondi:1962px, Sachs:1962wk}. The equivalence of these three seemingly disparate phenomena is an ``infrared triangle''~\cite{Strominger:2017zoo}. 

In the present work, we address the following question: is there a three-dimensional version of the infrared triangle?  Na\"{i}vely the answer is no; there are no propagating gravitons in three dimensions, and as such there are no soft graviton external states. Even so, we will find a suitable analogue as to be expected by the existence of this manuscript. 

The story goes as follows. It is already known~\cite{Ashtekar:1996cd,Barnich:2006av} that 3d gravity has an infinite-dimensional asymptotic symmetry in flat space, the 3d analogue of the BMS group often called BMS$_3$. This symmetry is generated by conserved charges at each angle at infinity, a ``supermomentum'' and a ``super-angular momentum.'' These charges label the configuration of boundary graviton degrees of freedom in flat space gravity. In the absence of matter, we use path integral techniques to reduce bulk 3d gravity to a quantum field theory description of these boundary degrees of freedom, which has BMS$_3$ as a global symmetry, building off of previous work on patches of flat space~\cite{Barnich:2017jgw,Merbis:2019wgk}.  (See~\cite{Bhattacharjee:2022pcb} for another attempt.) In fact, the BMS$_3$ symmetry is an infinite-dimensional extension of Carrollian symmetry, and this boundary description is an example of a Carrollian field theory of the magnetic type \cite{Henneaux:2021yzg,deBoer:2021jej}. There has been much recent interest in Carrollian field theories as a potential dual to flat space gravity; these theories are usually quite subtle quantum mechanically (see e.g.~\cite{deBoer:2023fnj,Cotler:2024xhb}). The boundary description we obtain, while a Carrollian theory, is not a holographic dual. Rather it is a mesoscopic description of 3d gravity, playing a similar role to the Schwarzian description of JT gravity~\cite{Jensen:2016pah,Maldacena:2016upp,Engelsoy:2016xyb} or the Alekseev-Shatashvili theory governing 3d gravity with negative cosmological constant~\cite{cotler2019theory}.

We carefully quantize this boundary theory, in this case a straightforward exercise without any subtleties like those uncovered in~\cite{Cotler:2024xhb}. This quantization reveals the existence of boundary graviton vacua labeled by elements of the space $\faktor{\text{Diff}(\mathbb{S}^1)}{PSL(2;\mathbb{R})}$. This result can be phrased quantum-mechanically:  there is a Hilbert space of soft graviton states $|P(\theta)\rangle$, eigenstates of the supermomentum $P(\theta)$ at each angle, and the 3d analogue of the soft graviton vacua of 4d gravity, with computable inner product along with a particularly simple $S$-matrix. 

Now consider gravity coupled to massless matter. Incoming and outgoing matter particles explicitly break the BMS$_3$ symmetry, but the non-conservation of the supermomentum and super-angular momentum are determined through Ward identities in terms of the ingoing/outgoing energy-momenta. These Ward identities are nothing more than the far-field limit of the constraint equations of gravity coupled to matter and play a central role in establishing an infrared triangle. The supertranslation Ward identity expresses that null energy injects (or removes) supermomentum, while null angular momentum injects (or removes) super-angular momentum. In the absence of in/outgoing matter, these Ward identities simply express the conservation of supermomentum and super-angular momentum, which together generate the BMS$_3$ asymptotic symmetry. The change in supermomentum required by in/outgoing energy is generated by a suitable large diffeomorphism, a superrotation, while changes in super-angular momentum are generated by supertranslations and superrotations. We corroborate these findings by a shockwave computation, in which we glue two halves of flat space across a shockwave, with the two halves related by a suitable large diffeomorphism. This diffeomorphism acts on the worldlines of particles in the far field leading to `rotational displacement' and `time displacement' memory effects. 

We further derive analogues of the gravitational soft and subleading soft theorem for scattering amplitudes. Asymptotic states are labeled both by quantum field theory degrees of freedom and the boundary graviton state, in such a way as to respect the Ward identities mentioned above. This determines the boundary graviton state at future timelike infinity in terms of the state at past timelike infinity and the in/outgoing distribution of null stress. There is a similar result for the change in super-angular momentum. 

The main results are the boxed equations in Section~\ref{sec:adding-matter}. The Ward identities are found in~\eqref{eq:boxedEOM}, the on-shell asymptotic symmetries in~\eqref{eq:boxedPoissonBrackets}, the soft theorems in~\eqref{E:boxedsoft1} and~\eqref{E:boxedsoft2}, and the memory effect in~\eqref{E:deviationboxed1}.

\begin{figure}[t]
  \centering
  \begin{tikzpicture}[scale=0.6]
    \node[align=center] (st) at (-4.5,0) {Soft\\Theorem\,};
    \node[align=center] (as) at (4.5,0)
    {Asymptotic\\Symmetries\,};
    \node[align=center] (me) at (0,7.8) {Memory\\Effects\,};
    \node[align=center] (fr) at (0,3.2) {\textbf{Ward} \\ \textbf{Identities}};

    \draw[thick,-] (st) -- (as);
    \draw[thick,-] (st) -- (me);
    \draw[thick,-] (as) -- (me);
    \draw[very thick, ->] (fr) -- (me);
    \draw[very thick, ->] (fr) -- (st);
    \draw[very thick, ->] (fr) -- (as);
  \end{tikzpicture}
  \caption{Three seemingly distinct features of 3d gravity coupled to massless matter -- infinite-dimensional asymptotic symmetry, memory effects, and a soft theorem -- all follow from the constraints of gravity coupled to matter in the far field. These constraints can be interpreted as sourced Ward identities at infinity. They generate the 3d gravity infrared triangle, which can also be regarded as an infrared trivalent vertex.}
  \label{fig:3d-triangle}
\end{figure}
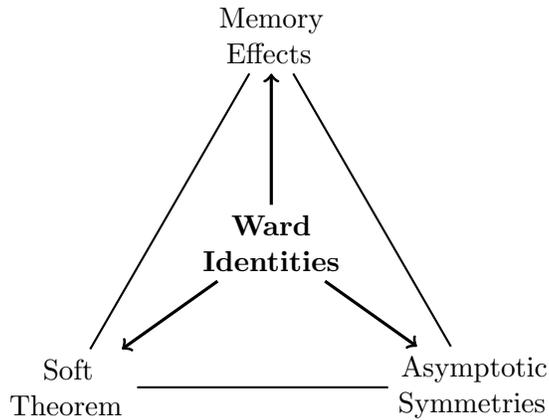

This analysis shows how infinite-dimensional asymptotic symmetries, memory effects, and soft theorems all follow from sourced Ward identities in the far field limit of 3d gravity coupled to matter. This completes the 3d gravity version of the infrared triangle, which perhaps should instead be thought of as an IR trivalent vertex. See Fig.~\ref{fig:3d-triangle}.  

As a byproduct of our pure gravity analysis we also resolve several extant puzzles about the partition function and Hilbert space interpretation of pure gravity in 3d flat space. For example, the Hilbert space is spanned by eigenstates of the supermomentum at every angle, all of which are energy eigenstates and so the $S$-matrix is diagonal in this basis. The full theory has a quadratic approximation whose Hilbert space includes the $\widehat{\text{BMS}}_3$ vacuum module of~\cite{Campoleoni:2015qrh,Campoleoni:2016vsh, Oblak:2016eij} as a subspace. The resolution of the full Hilbert space into sectors of definite angular momentum is unnecessarily difficult, leading to a deceptively complicated partition sum which obscures the simplicity of the $S$-matrix in the supermomentum basis. We anticipate that our findings as a whole will lead to further developments in 3d gravity, which is particularly tractable compared to its higher-dimensional counterparts. 

The remainder of the paper is organized as follows. In Section~\ref{sec:bdyQFT} we explain the quantization of 3d flat space quantum gravity in the absence of matter. After giving a review in Subsection~\ref{sec:review}, in Subsection~\ref{sec:geom-acti-bound} we rewrite 3d gravity in terms of a boundary theory described by a geometric action for boundary gravitons. We quantize this theory in Subsection~\ref{sec:hilb-space-ampl}, focusing on the Hilbert space, amplitudes, and symmetries of the theory. Then in Subsection~\ref{sec:partition-function} we use this theory to demystify the partition function of 3d gravity. In Section~\ref{sec:adding-matter} we add massless scalar matter and obtain an infrared triangle. We show in Subsection~\ref{subsec:Ward1} how constraints in the far field manifest as the Ward identities of an infinite-dimensional asymptotic symmetry. As an interesting aside, we show in Subsection~\ref{subsec:asympt1} how a minimally coupled massless scalar field coupled to gravity furnishes a novel representation of the BMS$_3$ group. We proceed in Section~\ref{subsec:soft1} by deriving the 3d soft graviton theorem, and memory effects in Section~\ref{sec:grav-memory}.  We conclude with a discussion in Section~\ref{sec:discussion}.   In Appendix~\ref{App:comments}, we discuss subtleties with $\delta$ distributions and Heaviside step functions $\theta$ apropos for our shockwave analysis.  In Appendix~\ref{App:absence}, we discuss an obstruction to obtaining a geometric action for $U(1)$ Chern-Simons theory coupled to a charged scalar. This serves as an illustration of how geometric actions fail to capture off-shell matter physics.

\section{Pure gravity}
\label{sec:bdyQFT}

In this Section we work out features of pure three-dimensional gravity with no cosmological constant, focusing on the quantum mechanical aspects of the theory including its Hilbert space and transition amplitudes.

\subsection{Review}
\label{sec:review}

Pure three-dimensional gravity in the second-order formulation is described by the action
\begin{equation}
	S = \frac{1}{16\pi G}\int d^3x \sqrt{-g} \,R + (\text{bdy})\,.
\end{equation}
Its asymptotically flat solutions are described by
\begin{equation}
\label{E:flatsolutions}
	ds^2 = -2 du dr + 16\pi G P(u,\theta) du^2  + 16\pi G J(u,\theta) du d\theta + r^2 d\theta^2\,,
\end{equation}
where Einstein's equations impose $P(u,\theta)=P(\theta), J(u,\theta)=j(\theta)+u P'(\theta)$.

Here $u$ is a Bondi time and future null infinity is reached as $r\to \infty$ with $(u,\theta)$ held fixed. The quantities $P(\theta)$ and $j(\theta)$ are the conserved supermomentum (aspect) and super-angular momentum (aspect), respectively. There is an infinite-dimensional asymptotic symmetry generated by the centrally extended version of the BMS$_3$ group, $\widehat{\text{BMS}}_3$~\cite{Ashtekar:1996cd,Barnich:2006av}.\footnote{In fact $\widehat{\text{BMS}}_3$ is the three-dimensional analogue of the extended BMS group in four dimensions~\cite{Barnich:2009se}, as well as the generalized BMS group in four dimensions\cite{Campiglia:2014yka}. The basic reason for this is that the group of conformal transformations of the celestial circle coincides with the diffeomorphism group of the celestial circle. The analogous statement is not true in higher dimension.} Infinitesimally these transformations act through vector fields
\begin{equation}
	\xi^{\mu}\partial_{\mu} = (T(\theta)+u Y'(\theta))\partial_u + \left( Y(\theta) - \frac{1}{r}(T'(\theta) + uY''(\theta))\right)\partial_{\theta} +\left( - rY'(\theta) +  T''(\theta) + u Y'''(\theta)\right)\partial_r + \hdots\,,
\end{equation}
where the dots indicate terms that are subleading at large $r$. Acting with this vector field on the metric~\eqref{E:flatsolutions} produces one with a change in $P$ and $j$ given by
\begin{align}
\begin{split}
\label{E:infinitesimalBMS}
	\delta_{\xi} P(\theta) & = 2Y'(\theta) P(\theta) +Y(\theta)P'(\theta) - \frac{1}{8 \pi G}Y'''(\theta) \,,
	\\
	\delta_{\xi} j(\theta) & = 2 Y'(\theta) j(\theta) + Y(\theta)j'(\theta) + 2T'(\theta) P(\theta) + T(\theta) P'(\theta) - \frac{1}{8\pi G}T'''(\theta)\,.
\end{split}
\end{align}
These transformations preserve the condition that the metric behaves near infinity as
\begin{equation}
\label{E:asymptoticallyflat1}
	ds^2 = - 2 \left(1 + O(1/r)\right) du dr + O(1) du^2 + O(1) du d\theta + r^2 d\theta^2\,.
\end{equation}
The finite form of the transformation is given by a reparameterization $\theta \to f(\theta)$ of the circle, a \emph{superrotation}, followed by a \emph{supertranslation} parameterized by $u \to u + \alpha(\theta)$, supplemented as needed so as to maintain the falloff condition~\eqref{E:asymptoticallyflat1}. The superrotation generates a non-trivial supermomentum, which is then left inert under a supertranslation, with the result
\begin{equation}
	P(\theta)  \to \widetilde{P}(\theta) = f'(\theta)^2 P(f(\theta)) - \frac{1}{8\pi G} \left\{ f(\theta),\theta\right\}\,,
\end{equation}
where $\{f(\theta),\theta\} = \frac{f'''(\theta)}{f'(\theta)} - \frac{3}{2}\left( \frac{f''(\theta)}{f'(\theta)}\right)^2$ is the Schwarzian derivative. In terms of the new supermomentum the super-angular momentum reads
\begin{equation}
	j(\theta) \to \widetilde{j}(\theta) =f'(\theta)^2 j(f(\theta)) + 2\alpha'(\theta) \widetilde{P}(\theta) + \alpha(\theta) \widetilde{P}'(\theta) - \frac{1}{8\pi G}\alpha'''(\theta)\,.
\end{equation}

Given a solution with some $(P(\theta),j(\theta))$ there is a continuous family of solutions related to it by $\widehat{\text{BMS}}_3$ transformations. The resulting space enjoys additional mathematical structure and is called a coadjoint orbit of $\widehat{\text{BMS}}_3$. For most orbits, one can use a $\widehat{\text{BMS}}_3$ transformation to set the supermomentum and super-angular momentum to constant values, $(P(\theta),j(\theta)) = (P,j)$. For pure Minkowski space those values are $(P_{\rm vac} = -\frac{1}{16\pi G},j_{\rm vac} = 0)$. The geometries with $P_{\rm vac}<P<0$ have conical singularities localized along the worldline of a massive particle, carrying spin if $j\neq 0$, while those with $P>0$ are ``flat space cosmologies'' with multiple boundaries~\cite{Cornalba:2003kd}. There are also orbits without a constant representative that we will not consider in the following; see~\cite{Oblak:2016eij} for a general discussion.

In the absence of matter, the supermomentum and super-angular momentum live on the orbit connected to pure Minkowski space, $\faktor{\text{BMS}_3}{ISO(2,1)}$. This orbit forms the phase space of pure 3d gravity with no cosmological constant. See~\cite{Oblak:2016eij} for discussion, and Section 2 of~\cite{cotler2019theory} for a primer to coadjoint orbits and their quantization. Any configuration on the orbit can be labeled by a reparameterization $f(\theta)$ and function $\alpha(\theta)$ by
\begin{align}
\begin{split}
	P(\theta)& = -\frac{1}{8\pi G} \left\{ \tan\left( \frac{f(\theta)}{2}\right),\theta\right\}\,,
	\\
	j(\theta) & = -\frac{1}{8\pi G}\left( 2\alpha'(\theta) \left\{\tan\left( \frac{f(\theta)}{2}\right),\theta\right\} + \alpha(\theta) \frac{d}{d\theta} \left\{ \tan\left( \frac{f(\theta)}{2}\right),\theta\right\}  + \alpha'''(\theta)\right)\,,
\end{split}
\end{align}
with $(f(\theta),\alpha(\theta))$ subject to an $ISO(2,1)$ equivalence, including
\begin{equation}
	\tan\left( \frac{f(\theta)}{2}\right) \sim \frac{a \tan\left( \frac{f(\theta)}{2}\right)+b}{c\tan\left( \frac{f(\theta)}{2}\right)+d}\,, \qquad ad-bc=1\,.
\end{equation}
We can think of this redundancy as that introduced by acting first on Minkowski space with a Poincar\'e transformation, then a $\widehat{\text{BMS}}_3$ transformation.

$\widehat{\text{BMS}}_3$ transformations are in fact generated by the supermomentum and super-angular momentum, with $P(\theta)$ generating supertranslations and $J(\theta)$ superrotations through the surface charge
\begin{equation}
	\mathcal{Q} = \frac{1}{2\pi} \int_0^{2\pi} d\theta\left( j(\theta)Y(\theta) + P(\theta)T(\theta)\right)\,.
      \end{equation}
In terms of the Fourier modes
\begin{equation}
	P_n = \frac{1}{2\pi} \int_0^{2\pi} d\theta\, e^{-in\theta}P(\theta)\,, \qquad j_n = \frac{1}{2\pi}\int_0^{2\pi} d\theta \,e^{-in\theta}j(\theta)\,,
\end{equation}
the algebra of charges is
\begin{align}
\begin{split}
\label{E:BMS3algebra}
	[P_n,P_m] & = 0\,,
	\\
	[P_n,j_m] & = (n-m)P_{n+m} + \frac{1}{8\pi  G}n^3\delta_{n,-m}\,,
	\\
	[j_n,j_m] & = (n-m)j_{n+m} \,.
\end{split}
\end{align}

So far we have focused on the charges near future null infinity. Let us call them $(P_+,j_+)$. Near past null infinity, the supermomentum and super-angular momentum are given by 
\begin{equation}
\label{E:antipodal}
	P_-(\theta) = P_+(\theta+\pi)\,, \qquad j_-(\theta) =  j_+(\theta+\pi)\,,
\end{equation}
which is an analogue of antipodal matching in three dimensions \cite{Prohazka:2017equ,Compere:2017knf}.

\subsection{Boundary action}
\label{sec:geom-acti-bound}

In this work, we are interested in three-dimensional gravity on spaces with the topology of Minkowski space. In that context, pure gravity can be reduced to a boundary system, much like with negative~\cite{cotler2019theory} or positive~\cite{Cotler:2019nbi} cosmological constant. Such a reduction was performed in a patch of spacetime including future null infinity~\cite{Merbis:2019wgk}. The basic idea is as follows. One starts from the first-order formulation of pure 3d gravity, which can locally be recast as a Chern-Simons gauge theory with gauge group $ISO(2,1)$. This gauge theory formulation should always be regarded as a mnemonic rather than a literal definition of 3d gravity; its ``connection'' is a linear combination of the dreibein and spin connection, which one integrates over upon quotienting by diffeomorphisms and local Lorentz transformations, which at the infinitesimal level and around saddle points (like 3d flat space) are equivalent to infinitesimal $ISO(2,1)$ gauge transformations. Even so one can borrow methods from the canonical quantization of Chern-Simons theory to proceed. 

In this particular quantization problem we are interested in fluctuations around flat Minkowski space. Upon choosing a foliation for time, the constant time slice is a topological disk, and the quantization resembles that of Chern-Simons theory on a cylinder. The temporal components of the dreibein and spin connection appear linearly in the gravity action and can be integrated out, enforcing constraints that can be recast as the flatness of the spatial ``connection'' on the disk. These constraints can be solved leading to a residual description in terms of an $ISO(2,1)$-valued field on the boundary, with no moduli, and after enforcing the boundary conditions of 3d flat space gravity one lands on a description in terms of two boundary degrees of freedom, what we call $\alpha$ and $f$, as they will bear a very close connection with the $\widehat{\text{BMS}}_3$ transformations parameterized by $\alpha$ and $f$ described in the last Subsection.

The authors of~\cite{Merbis:2019wgk} carried out this procedure for the future half of Minkowski space so that the boundary in question is future null infinity. In terms of null boundary time $u$ and the angle $\theta$ on the celestial circle, the resulting action they found can be rewritten as
\begin{equation}
\label{E:action1}
	S_+[\alpha,f] = -\int du d\theta \left( \alpha \partial_u P + P\right)\,, \qquad P = - \frac{1}{8\pi G}\left\{ \tan\left( \frac{f}{2}\right),\theta\right\}\,.
\end{equation}
At constant time $u$ the fields $\alpha$ and $f$ are constrained: they parameterize the vacuum coadjoint orbit of $\widehat{\text{BMS}}_3$, $\faktor{\text{BMS}_3}{ISO(2,1)}$, meaning that they are subject to the boundary conditions
\begin{equation}
	f(\theta+2\pi,u) = f(\theta,u) + 2\pi\,, \qquad \partial_{\theta} f >0\,, \qquad \alpha(\theta+2\pi,u) = \alpha(\theta,u)\,,
\end{equation}
and a time-dependent $ISO(2,1)$ gauge symmetry
\begin{align}
\begin{split}
\label{E:gauge}
	\tan \left( \frac{f(\theta,u)}{2}\right) & \sim \frac{a(u) \tan\left( \frac{f(\theta,u)}{2}\right)+b(u)}{c(u)\tan\left( \frac{f(\theta,u)}{2}\right)+d(u)}\,, \qquad ad-bc = 1\,,
	\\
	\alpha(\theta,u) & \sim  \alpha(\theta,u) + \frac{a_0(u) +a_1(u) e^{i f(\theta,u)} + a_{-1}(u) e^{-i f(\theta,u)} }{\partial_{\theta} f(\theta,u)}\,,
\end{split}
\end{align}
with $(a,b,c,d;a_{-1,0,1})$ labeling an element of $ISO(2,1) = PSL(2;\mathbb{R})\ltimes \mathbb{R}^3=SO(2,1)^{+}\ltimes \mathbb{R}^3$. $(a,b,c,d)$ labels a $PSL(2;\mathbb{R})$ subgroup and $a_{-1,0,1}$ a $\mathbb{R}^3$ subgroup. These peculiar properties imply that at constant time $u$, $f(\theta)$ is a reparameterization of the (celestial) circle subject to a $PSL(2;\mathbb{R})$ redundancy. That is, it is an element of the quotient space $\faktor{\text{Diff}(\mathbb{S}^1)}{PSL(2;\mathbb{R})}$, which has made a frequent appearance in JT and 3d gravity. 

The Schwarzian derivative is $PSL(2;\mathbb{R})$-invariant, and so the action is manifestly invariant under the $PSL(2;\mathbb{R})$ subgroup of $ISO(2,1)$ corresponding to $(a,b,c,d)$. Its invariance under $a_{-1,0,1}$ is a bit more non-trivial to work out, although we give a simple demonstration below when realizing this model as a large radius limit of AdS$_3$ gravity.

The approach we just outlined is the ``constrain first'' approach to quantization. In general, it should be understood as an approximation, sometimes exact, to the full path integral. What it neglects is the integral over the gauge-fixing ghosts, which depending on the choice of gauge, can contribute a ghost bilinear to the constraints imposed after integrating out the $u$-components of the dreibein and spin connection.
For a Chern-Simons theory with compact gauge group and in Coulomb gauge, the constraint is precisely the classical one and with some effort, one can show that the full functional integral collapses to one over the moduli space of flat spatial connections. In other words, the constrain first approach gives the exact answer in that setting. However in 3d gravity, there does not appear to be an analogue of Coulomb gauge: we have been unable to find a gauge in which the Faddeev-Popov ghosts decouple from the temporal component of the dreibein, and so the ghosts contribute to the constraint. This can in principle generate an effectively two-loop correction to the quantization obtained by the constrain first procedure, although in the remainder of this Section, we will see that there is good reason to think that no such correction is generated.

The ensuing boundary path integral is then
\begin{equation}
\label{E:futureZ}
	\int \frac{[d\alpha][df]}{ISO(2,1)}e^{i S_+}\,, 
\end{equation}
where we have indicated the time-dependent $ISO(2,1)$ quotient in the measure. The patch of Minkowski space we have just discussed is the future half. There is a similar reduction to a boundary system for the past half, along with gluing conditions that match the two. The full description for pure 3d gravity in all of asymptotically flat space reads
\begin{equation}
\label{E:totalZ}
	Z = \int \frac{[d\alpha_+][df_+]}{ISO(2,1)} \frac{[d\alpha_-][df_-]}{ISO(2,1)} e^{i (S_+[\alpha_+,f_+] + S_-[\alpha_-,f_-])}\,, 
\end{equation}
with
\begin{align}
\begin{split}
\label{E:totalS}
	S_+ &= -\int du d\theta \left( \alpha_+ \partial_u P_+ + P_+\right)\,, 
	\\
	S_- & =- \int dv d\theta\left( \alpha_- \partial_v P_- + P_-\right)\,,
	\\
	P_{\pm}& = -\frac{1}{8\pi G}\left\{ \tan\left( \frac{f_{\pm}}{2}\right),\theta\right\}\,.
\end{split}
\end{align}
The fields $(\alpha_+,f_+)$ and $(\alpha_-,f_-)$ are, at constant time, elements of the vacuum coadjoint orbit of BMS$_3$ indicated above, and thus subject to independent $ISO(2,1)$ gauge symmetries. The limit $u\to -\infty$, and $v\to +\infty$ corresponds to spatial infinity, as reached from future and past null infinity respectively. The fields are joined together there with the boundary conditions
\begin{equation}
\label{E:gluing}
	\lim_{u\to -\infty} f_+(\theta,u) = \lim_{v\to \infty} f_-(\theta+\pi,v)\,, \qquad \lim_{u\to -\infty} \alpha_+(\theta,u) =  \lim_{v\to\infty}\alpha_-(\theta+\pi,v)\,,
\end{equation}
along with a suitable gluing of the two $ISO(2,1)$ gauge symmetries.

The basic gauge-invariant operators in this theory are the supermomenta $P_{\pm}$, the super-angular momentum $J_{\pm}$ defined below in~\eqref{E:J}, and composites built from them. These are in fact the only gauge-invariant local operators of the model. There are other bilocal operators as is familiar from the Schwarzian description of JT gravity or the Alekseev-Shatashvili description of pure three-dimensional AdS gravity. These operators are characterized by two insertions at the same time and are given by~\cite{Merbis:2019wgk}
\begin{equation}
    \mathcal{B}_+(u,\theta_1,\theta_2;\Delta) = \left( \frac{\partial_{\theta_1}f_+(u,\theta_1) \partial_{\theta_2} f_+(u,\theta_2)}{4\sin^2\left( \frac{f_+(u,\theta_1)-f_+(u,\theta_2)}{2}\right)}\right)^{\Delta}\,,
\end{equation}
with a similar expression for a past bilocal. In pure AdS$_3$ gravity, the two-point function of bilocal operators computes vacuum Virasoro blocks. One can perform similar computations here with a trivial result, as is clear from the transition amplitudes of the model summarized in~\eqref{E:Amplitude1}. It would be interesting to understand the relation between this result and the work of~\cite{Hijano:2019qmi}.

\subsection{Recovering classical gravity}

Several comments are in order to unpack the physics in~\eqref{E:action1}. The fields appearing therein are the boundary graviton degrees of freedom. They parameterize fluctuations of the metric, i.e.\ correspond to metric fluctuations at the indicated powers of the large radius expansion in~\eqref{E:asymptoticallyflat1}, and as such we integrate over them subject to initial and final boundary conditions. As we mentioned, at fixed time $(\alpha,f)$ describe an element of the quotient space $\faktor{\text{BMS}_3}{ISO(2,1)}$.

This fact is worth mentioning for two reasons. First, this is precisely the phase space of pure 3d gravity with zero cosmological constant. Since we integrate over $(\alpha,f)$ promoted to functions of time, we see that we are dealing with a phase space path integral. Second, this space is a coadjoint orbit of $\widehat{\text{BMS}}_3$, the centrally extended version of BMS$_3$. Coadjoint orbits are particularly ``nice'' symplectic spaces, endowed with a group action (in this case under $\widehat{\text{BMS}}_3$) and an invariant symplectic form. This structure is what will easily allow us to perform the path integral over $(\alpha,f)$, although we note that one does not need to be versed in the coadjoint orbit machinery to proceed.

Before treating the model quantum mechanically, let us see how to recover known facts about classical 3d gravity from this boundary graviton description. The equations of motion of the model, which follow from varying $\alpha$ and $f$, and \emph{not} $P$ (since $P$ is a constrained field), read
\begin{align}
\begin{split}
\label{E:flatEoMs}
	0 & = \partial_u P\,,
	\\
	0 & = \partial_u J -\partial_{\theta} P-2\partial_{\theta} \alpha \partial_u P-\alpha \partial_{\theta}\partial_u P\,,
\end{split}
\end{align}
where
\begin{equation}
\label{E:J}
	J =   \alpha \partial_{\theta} P  + 2 P \partial_{\theta} \alpha- \frac{1}{8 \pi G} \partial_{\theta}^3\alpha\,.
\end{equation}
On-shell these take the form of conservation equations, $\partial_u P = 0$ and $\partial_u J = \partial_{\theta} P$, so that $P = P(\theta)$ and $J = j(\theta) + u P'(\theta)$. In terms of the original fields $\alpha$ and $f$, this corresponds to a time-independent profile for $f$, and a profile for $\alpha$ which is linear in time. 

We then see that $j$ and $P$ are conserved. They are also invariant under $ISO(2,1)$ gauge transformations. Physically they are the super-angular momentum and supermomentum carried by the spacetime. One can reconstruct the corresponding 3d metric, which for a particular choice of coordinates is precisely that presented in~\eqref{E:flatsolutions}.

The conservation of $j$ and $P$ is precisely the consequence of the $\widehat{\text{BMS}}_3$ asymptotic symmetry of pure 3d gravity, generated by superrotations and supertranslations. We can identify this symmetry directly in the boundary graviton description. Given an on-shell configuration of $(\alpha,f)$, i.e.\ $J = j(\theta)+u P'(\theta)$ and $P=P(\theta)$, which by the gauge symmetry we can take to be of the form $f=f(\theta)$ and $\alpha =\bar{ \alpha}(\theta) + u $, these symmetries act as
\begin{align}
\begin{split}
	\delta f & = f'(\theta) Y(\theta)\,,
	\\
	\delta \alpha & = T(\theta)\,,
\end{split}
\end{align}
which by Eqs.~\eqref{E:action1} and~\eqref{E:J} induce precisely the infinitesimal variations of $(P,j)$ given in~\eqref{E:infinitesimalBMS}. Using the Poisson brackets implied by the form of the boundary action~\eqref{E:action1},
\begin{equation}
	\omega = - \int_0^{2\pi} d\theta\, d\alpha \wedge dP\,,
\end{equation}
it is straightforward to compute the algebra of $\widehat{\text{BMS}}_3$ charges~\eqref{E:BMS3algebra}. Meanwhile, the gluing conditions~\eqref{E:gluing} imply that the supermomentum and super-angular momentum obey the antipodal matching conditions~\eqref{E:antipodal}. 

\subsection{Hilbert space and transition amplitudes}
\label{sec:hilb-space-ampl}

Now we endeavor to quantize the boundary description obtained above. We begin with a quantization of the future half of flat space corresponding to~\eqref{E:futureZ}. We do so first in a heuristic way, using a canonical quantization of the quadratic approximation to the full theory. Then, we give an all-orders path integral description. The past half can be treated in the same way. We patch the results together in Subsection~\ref{S:futureAndPast}.

 \subsubsection{A quantum mechanical treatment, take 1}
 
The action
 \begin{equation*}
 	S_+ =- \int dud\theta \left( \alpha \partial_u P + P\right)\,, \qquad P = -\frac{1}{8\pi G}\left\{ \tan\left( \frac{f}{2}\right),\theta\right\}\,,
 \end{equation*}
is in Hamiltonian form, where, roughly speaking, $\alpha$ is the momentum conjugate to $-P$ and the Hamiltonian is $H = \int_0^{2\pi} d\theta\, P$, i.e.\ the Fourier zero-mode of $P$. This is not exactly correct, as $P$ is a constrained field and $\alpha$ is subject to a gauge symmetry~\eqref{E:gauge}. However, as we will see, this is a useful heuristic for interpreting this theory. 

As a first attempt at quantization, let us take a saddle point approximation at small $G$ around the configuration of minimum energy. The Hamiltonian is bounded below by
\begin{equation}
	H \geq -\frac{1}{8 G}\,,
\end{equation}
where the minimum value is achieved at $f = \theta$ (up to a $PSL(2;\mathbb{R})$ gauge transformation). Expanding around this point,
\begin{equation}
	f(\theta,u) = \theta + \varepsilon(\theta,u)\,, \qquad P = -\frac{1}{16\pi G} -\frac{1}{8\pi G}(\varepsilon'''-\varepsilon') + O(\varepsilon^2)\,,
\end{equation}
the quadratic approximation to the action for $(f,\alpha)$ reads
\begin{equation}
	S_+ =S_0 +\frac{1}{8\pi G}\int d\theta du\left( \alpha \partial_u \!\left( \varepsilon''' - \varepsilon'\right) -\frac{1}{2}\left( \varepsilon''^2- \varepsilon'^2\right)\right) + \hdots\,,
\end{equation}
where $S_0 =\int du \frac{1}{8G}$ is the saddle-point value of the action and the dots indicate terms with at least three powers of fields. Using~\eqref{E:gauge} we see that the fluctuations $(\varepsilon,\alpha)$ are subject to the gauge redundancy
\begin{align}
\begin{split}
	\varepsilon(\theta,u)& \sim \varepsilon(\theta,u) + \tilde{a}_{1}(u) e^{i \theta} + \tilde{a}_0(u) +\tilde{a}_{-1}(u)e^{-i\theta}\,,
	\\
	\alpha(\theta,u) & \sim \alpha(\theta,u) + a_1(u) e^{i\theta} + a_0(u) + a_{-1}(u) e^{-i\theta}\,,
\end{split}
\end{align}
Fourier transforming in angle,
\begin{align}
\begin{split}
	\alpha & = \sum_n \alpha_n(u) e^{i n \theta}\,,
	\\
	\varepsilon & = \sum_n \varepsilon_n(u) e^{in\theta}\,,
\end{split}
\end{align}
we can gauge-fix
\begin{equation}
	\varepsilon_{-1,0,1}(u) = \alpha_{-1,0,1}(u) = 0\,.
\end{equation}
This means that the supermomentum $P$, to linearized order in fluctuations, receives contributions only from $|n|>1$ modes, with
\begin{equation}
\label{E:Pmodes}
	P = -\frac{1}{16\pi G} + \sum_{|n|>1} P_n(u) \,e^{in\theta}\ + \cdots\,, \qquad P_n = \frac{i}{8\pi G} n(n^2-1) \varepsilon_n\,.
\end{equation}
We then have
\begin{equation}
	S_+ = S_0 + \frac{1}{4G}\int du \sum_{|n|>1} \left( -i n(n^2-1)\alpha_{-n} \,\partial_u \varepsilon_n  - \frac{1}{2}n^2(n^2-1)|\varepsilon_n|^2\right) + \hdots\,.
\end{equation}
which when rewritten in terms of the $P_n$'s becomes 
\begin{equation}
	S_+ = S_0 - \int du \sum_{|n|>1} \left(  \alpha_{-n} \partial_u P_n + \frac{8\pi^2 G}{n^2-1} |P_n|^2\right) + \hdots\,.
\end{equation}

We can now canonically quantize the quadratic approximation. Promoting the $(P_n,\alpha_m)$ to operators $(\hat{P}_n,\hat{\alpha}_m)$, we assign the canonical commutation relations and identify the Hamiltonian
\begin{equation}
\label{E:canonicalQuant}
	[\hat{\alpha}_n,\hat{P}_m] = i \delta_{n,-m}\,, \qquad \hat{H} = -\frac{1}{8G}+ 8\pi^2 G \sum_{|n|>1} \frac{\hat{P}_n \hat{P}_{-n}}{n^2-1}\,.
\end{equation}
The weak coupling limit is that of small $G$, so that the quadratic part of the Hamiltonian is perturbatively small. The $\hat{P}_n$ can be simultaneously diagonalized, giving a basis for the boundary graviton Hilbert space given by $\{ |\{P_n\}\rangle\}$. This perturbative Hilbert space has a ``vacuum state'' $|\{P_n=0\}\rangle$ of energy $E =-\frac{1}{8G}$. Now consider the subspace of states given by the vacuum together with the states obtained by acting on it with the $J_n$. That subspace is isomorphic to the vacuum $\widehat{\text{BMS}}_3$ module of~\cite{Campoleoni:2015qrh,Campoleoni:2016vsh, Oblak:2016eij}. However, the perturbative Hilbert space spanned by the $|\{P_n\}\rangle$ is clearly much bigger. All of these states are eigenstates of the Hamiltonian, and we can assign them the inner product
\begin{equation}
\label{E:innerProduct1}
	\langle \{P'_n\}|\{P_m\}\rangle = \prod_{|n|>1} \delta(P'_n-P_n)=\delta[P'-P] \,,
\end{equation}
where the last equation defines (for now) what we mean by a delta function on the supermomentum. Recalling that $P_n$ is the $n$th Fourier mode of $P$ around the circle, we see that states are effectively labeled by their supermomentum $P(\theta)$, and we can define the supermomentum operator
\begin{equation}
	\hat{P}(\theta) = -\frac{1}{16\pi G} + \sum_{|n|>1} \hat{P}_n e^{i n \theta}\,,
\end{equation}
so that the  $|\{P_n\}\rangle$ are supermomentum eigenstates with eigenvalue $P(\theta)=-\frac{1}{16\pi G} +  \sum_{|n|>1}P_n e^{i n \theta}$. To linearized order in fluctuations, these are the supermomentum profiles consistent with $P$ obeying the constraint $P = -\frac{1}{8\pi G}\left\{ \tan\left( \frac{f}{2}\right),\theta\right\}$ with $f$ an element of $\faktor{\text{Diff}(\mathbb{S}^1)}{PSL(2;\mathbb{R})}$. These supermomentum eigenstates obey a delta-function norm which, at least for $P$ near $-\frac{1}{16\pi G}$, is given by~\eqref{E:innerProduct1}. Under evolution by time $T$, we then have the transition amplitudes of the quadratic theory
\begin{equation}
	\langle \{ P'_n\}| e^{-i \hat{H}T}|\{P_m\}\rangle_{\rm quadratic} =  e^{-i E[P]T}  \delta[P'-P]\,, \quad E[P] = -\frac{1}{8G} +8\pi^2 G\sum_{|n|>1} \frac{|P_n|^2}{n^2-1} \,.
\end{equation}

We may also consider $J$ in~\eqref{E:J}. To linearized order in fluctuations we have
\begin{equation}
	J = -\frac{1}{8\pi G} (\alpha'''-\alpha') + \hdots\,,
\end{equation}
so that its Fourier modes are
\begin{equation}
\label{E:Jn}
	J_n = \int_0^{2\pi} \frac{d\theta}{2\pi}\, e^{-i n \theta} J = \frac{i}{8\pi  G}\sum_{|n|>1} n(n^2-1)\alpha_n + \cdots\,.
\end{equation}
These can be promoted to operators $\hat{J}_{-n}$, which thanks to the commutator~\eqref{E:canonicalQuant} acts as $\frac{n(n^2-1)}{8\pi G} \frac{\partial}{\partial P_n}$ on the wavefunction in the supermomentum basis at time $u=0$. One can also evaluate the Heisenberg equation of motion for $\hat{J}_n$, with the result
\begin{equation}
	\frac{d\hat{J}_n}{du} = in\hat{P}_n\,,
\end{equation}
which is what we expected from the on-shell conservation equation $\partial_u J = \partial_{\theta} P$. We then define the modes of the super-angular momentum, 
\begin{equation}
	\hat{j}_n = \hat{J}_n - i u n \hat{P}_n\,,
\end{equation}
which are clearly conserved, and whose matrix elements are given by
\begin{equation}
	\langle \{P'_{n'}\}| \hat{j}_{-m} |\{P_n\}\rangle = \frac{m(m^2-1)}{8\pi G} \frac{\partial}{\partial P_m} \delta[P'-P]\,.
\end{equation}

What are the lessons so far in this quadratic approximation?
\begin{enumerate}
	\item The Hilbert space of boundary gravitons can be labeled by eigenkets of definite supermomentum $P(\theta)$ obeying the constraint~\eqref{E:action1}.
	\item The supermomentum and super-angular momentum are conserved mode by mode, although only the conservation of supermomentum is manifest.
\end{enumerate}

The constraint on $P$ and gauge redundancy of $\alpha$ prohibit us from performing a non-perturbative version of the canonical quantization. However, next, we simply perform the path integral over $(f,\alpha)$ to obtain the full Hilbert space and amplitudes. These results are morally the same as those obtained in the quadratic approximation.

\subsubsection{A quantum mechanical treatment, take 2}
 
 Now we endeavor to exactly compute the path integral
 \begin{equation*}
 	\int \frac{[d\alpha][df]}{ISO(2,1)} \,e^{i S_+}\,, \qquad S_+ = -\int dud\theta \left( \alpha \partial_u P + P\right)\,.
 \end{equation*}
 We consider a finite time interval $T$. Writing the action above, where the time derivative acts on $P$, we see that we have a consistent variational principle if we fix $P$ on the initial and final time slices. We cannot pick arbitrary initial and final $P$. These boundary conditions must satisfy the constraint that $P = -\frac{1}{8\pi G}\left\{ \tan\left( \frac{f}{2}\right),\theta\right\}$ with $f$ an element of $\faktor{\text{Diff}(\mathbb{S}^1)}{PSL(2;\mathbb{R})}$. Let the initial profile for $P$ be $P_1(\theta)$, and the final profile be $P_2(\theta)$. Already then this gives us our first result. In the finite-time path integral, the initial and final boundary conditions prepare the initial and final states. Here, we see that we can label states by definite supermomentum, just as in the quadratic approximation. 
 
 Call the initial state $|P_1(\theta)\rangle$ and the final state $|P_2(\theta)\rangle$. The field $\alpha$ acts as a Lagrange multiplier, enforcing that $\partial_u P = 0$, i.e.\ $P$ is conserved, so that the finite-time amplitude will be delta-localized to $P_2(\theta)=P_1(\theta)$. So, morally, the finite-time amplitude is
\begin{equation}
	\langle P_2(\theta)| \widehat{\mathcal{U}}|P_1(\theta)\rangle = \delta[P_2-P_1] e^{-i E[P_1]T}\,.
\end{equation}
We would like to know the precise definition of the delta function and the energy of the supermomentum eigenstate.
  
Because the amplitude is supported only at $P_1 = P_2$, we can take $P_1$ and $P_2$ to be perturbatively close to each other. Furthermore, let us time-slice the path integral, and look only over a single time interval of time $\epsilon$. Let $P_1$ denote the initial profile of $P$ at $u=0$, and, with an abuse of notation, $P_2$ the profile at $u=\epsilon$. Over the single time step, we integrate over $\alpha$ while $P$ is fixed by the boundary conditions to be
\begin{equation}
	P(\theta,u) = P_1(\theta) + (P_2(\theta)-P_1(\theta))\frac{u}{\epsilon}\,.
\end{equation}
Approximating the Hamiltonian term in the action halfway between the initial and final times,
\begin{equation}
	\int d\theta du \,P \to \epsilon\int_0^{2\pi} d\theta\frac{P_1(\theta)+P_2(\theta)}{2}\,,
\end{equation}
the short-time amplitude is only an integral over $\alpha$ at this time,
\begin{equation}
	\langle P_2(\theta)|\widehat{\mathcal{U}}(\epsilon)|P_1(\theta)\rangle = \int \frac{[d\alpha(\theta)]}{\text{gauge}} \exp\left( -i \int_0^{2\pi} d\theta \left( \alpha(\theta) (P_2(\theta)-P_1(\theta)) + \epsilon\frac{P_1(\theta)+P_2(\theta)}{2}\right)\right)\,,
\end{equation}
and we integrate over $\alpha$ modulo the gauge redundancy. 

To determine the redundancy, let us first parameterize $P_1\approx P_2$ in a useful way:
\begin{align}
\begin{split}
\label{E:dP}
	P_1(\theta) &= - \frac{1}{8\pi G}\left\{ \tan\left( \frac{f(\theta)}{2}\right),\theta\right\} \,, 
	\\
	\qquad P_2(\theta) &= -\frac{1}{8\pi G}\left\{ \tan\left( \frac{f(\theta+ Y(\theta))}{2}\right) ,\theta\right\}
	\\
	& = P_1(\theta) + Y(\theta)P_1'(\theta) + 2 P_1(\theta)Y'(\theta) - \frac{1}{8\pi G}Y'''(\theta) + O(Y^2)\,,
\end{split}
\end{align}
where the $O(Y)$ part of the last line is the transformation of $P_1$ in~\eqref{E:infinitesimalBMS}. Meanwhile, the gauge symmetry acts on $\alpha$ as
\begin{equation}
	\alpha (\theta) \sim \alpha(\theta) + \frac{a_1 e^{i f(\theta)}+a_0 + a_{-1}e^{-if(\theta)}}{f'(\theta)} + O(Y)\,.
\end{equation}
Now let us exploit the fact that $f(\theta)$ is a reparameterization of the circle, and so we may trade $\theta$ for $f$. Expanding $\alpha$ and $Y$ in Fourier modes of $f$,
\begin{align}
\begin{split}
	\alpha(\theta) & = \frac{1}{f'(\theta)} \sum_n \alpha_n e^{inf(\theta)}\,,
	\\
	Y(\theta) & = \frac{1}{f'(\theta)} \sum_n Y_n e^{inf(\theta)}\,,
\end{split}
\end{align}
the part of the short-time action that depends on $\alpha$ becomes
\begin{equation}
\label{E:shortTimeAlpha}
	\int_0^{2\pi} d\theta\, \alpha(\theta)(P_2(\theta)-P_1(\theta)) = - \frac{i}{4G} \sum_n n(n^2-1) \alpha_n Y_{-n} + O(Y^2)\,.
\end{equation}
The $ISO(2,1)$ gauge redundancy allows us to fix $\alpha_{n=-1,0,1}=Y_{n=-1,0,1} = 0$. 

Note that this part of the short-time action is independent of the initial supermomentum represented through $f$. This is a consequence of the fact that the boundary graviton action is a quantization of a coadjoint orbit. This part of the action is fixed by the presymplectic potential of the coadjoint orbit, here $\faktor{\text{BMS}_3}{ISO(2,1)}$, which is invariant under BMS$_3$ action. In practice, this means that the symplectic structure is, in the right variables, the same everywhere in field space, which is of course what we find in~\eqref{E:shortTimeAlpha}.

With a choice of normalization, the integration measure over $\alpha$ is then\footnote{The full measure over $\alpha$ and $Y$ at a single time step is fixed by the symplectic structure to be
\begin{equation}
	\frac{[d\alpha(\theta)][dY(\theta)]}{ISO(2,1)} \propto \prod_{|n|>1} n(n^2-1)d\alpha_n dY_{-n} (1+O(Y))
\end{equation}
We define a measure over the $\alpha_n$'s to not include factors of $n(n^2-1)$, so that the measure over $Y_n$'s does include such a factor. Expressed in terms of supermomentum fluctuations $dP_n$, this means that the measure over $dP_n$ is simply $\prod_{|n|>1} dP_n$, commensurate with the delta function in~\eqref{E:deltaPfinal}.
}
\begin{equation}
	\frac{[d\alpha(\theta)]}{\text{gauge}} = \prod_{|n|>1} d\alpha_n \left( 1+ O(Y)\right)\,,
\end{equation}
so that the integral over $\alpha$ then gives
\begin{equation}
	\langle P_2(\theta)|\,\widehat{\mathcal{U}}(\epsilon)|P_1(\theta)\rangle =e^{-i E \epsilon} \prod_{|n|>1} \frac{8\pi G}{|n|(n^2-1)} \delta(Y_n)\,, \qquad E = \int_0^{2\pi} d\theta \frac{P_1(\theta)+P_2(\theta)}{2}\,.
\end{equation}
This defines the delta function on supermomentum to be
\begin{equation}
	\delta[P_2-P_1] = \prod_{|n|>1} \frac{8 \pi G \delta(Y_n)}{|n|(n^2-1)}\,,
\end{equation}
which notably does not depend on contributions of $O(Y^2)$ and higher in the short-time action, as is usual for the Jacobian of a functional delta function.

Equivalently, let $dP_n$ denote the $n$th Fourier mode of $f'(\theta)(P_2-P_1)$ in the $f$ coordinate. Using~\eqref{E:dP} we find
\begin{equation}
	dP_n = \int_0^{2\pi} \frac{df}{2\pi} \,e^{-i n f}f'(\theta)(P_2(\theta)-P_1(\theta)) = \frac{i}{8\pi G}n(n^2-1)Y_n\,,
\end{equation}
so that the delta function can be instead written as 
\begin{equation}
\label{E:deltaPfinal}
	\delta[P_2-P_1] = \prod_{|n|>1} \delta(dP_n)\,.
\end{equation}
 
Building up the finite-time amplitude out of the short-time one, we then find
\begin{equation}
\label{E:Amplitude1}
	\langle P_2(\theta)| \,\widehat{\mathcal{U}}(T)|P_1(\theta)\rangle = \delta[P_2-P_1] e^{-i E[P_1]T}\,, \qquad E[P] = \int_0^{2\pi} d\theta\, P(\theta)\,,
\end{equation}
as expected. The short-time limit of the amplitude also gives the inner product of supermomentum eigenstates,
\begin{equation}
	\langle P_2(\theta)|P_1(\theta)\rangle = \delta[P_2-P_1]\,.
\end{equation}

We then find results that are physically the same as in the canonical quantization of the quadratic theory above. Supermomentum eigenstates give a basis of the boundary graviton Hilbert space, and supermomentum is conserved mode by mode. 

\subsubsection{Aside: flat space limit of AdS$_3$}
\label{sec:limit}

One way to understand pure 3d gravity in flat space is as arising from a large radius limit of pure 3d gravity on global AdS$_3$ \cite{Barnich:2012aw,Barnich:2012rz}. In~\cite{cotler2019theory} two of us obtained a boundary graviton description of the latter. In this Subsection, we work out this relation.

Pure 3d gravity with negative cosmological constant on the global AdS$_3$ cylinder can be reduced to a boundary system in terms of chiral boundary degrees of freedom $\phi(t,\theta), \overline{\phi}(t,\theta)$ as
\begin{equation}
\label{E:alekseevShatashvili}
S = - \frac{C}{24\pi} \int dtd\theta  \left(\frac{\phi'' \partial_+ \phi'}{\phi'^2} - \phi' \partial_+ \phi + \frac{\overline{\phi}'' \partial_- \overline{\phi}'}{\overline{\phi}'^2} - \overline{\phi}' \partial_- \overline{\phi} \right)\,,
\end{equation}
where $\theta \sim \theta + 2\pi$, $\partial_{\pm} = \frac{1}{2}(\partial_\theta \pm \partial_t)$, the prime $'$ denotes a spatial derivative, and $C = \frac{3L}{2G}$ is the classical Brown-Henneaux central charge. The fields $\phi, \overline{\phi}$ are at fixed time, reparameterizations of the angular circle
\begin{align*}
\phi(t,\theta + 2\pi) &= \phi(t,\theta) + 2\pi\,, \qquad \phi' > 0 \\
\overline{\phi}(t, \theta + 2\pi) &= \overline{\phi}(t,\theta) + 2\pi\,, \qquad \overline{\phi}' > 0\,.
\end{align*}
and, at fixed time, are elements of the quotient space $\faktor{\text{Diff}(\mathbb{S}^1)}{PSL(2;\mathbb{R})}$:
\begin{equation}
\label{E:AdSgauge}
\tan\left(\frac{\phi(t,\theta)}{2}\right) \sim \frac{a(t) \, \tan\left(\frac{\overline{\phi}(t,\theta)}{2}\right) + b(t)}{c(t) \tan\left(\frac{\overline{\phi}(t,\theta)}{2}\right) + d(t)}\,, \quad \tan\left(\frac{\overline{\phi}(t,\theta)}{2}\right) \sim \frac{\overline{a}(t) \, \tan\left(\frac{\overline{\phi}(t,\theta)}{2}\right) + \overline{b}(t)}{\overline{c}(t) \tan\left(\frac{\overline{\phi}(t,\theta)}{2}\right) + \overline{d}(t)}\,,
\end{equation}
where $ad-bc=\bar{a}\bar{d}-\bar{b}\bar{c}=1$. The fields $\phi, \overline{\phi}$ parameterize subleading fluctuations of the AdS$_3$ metric near the boundary, in particular, they correspond to metric fluctuations that respect the asymptotically AdS boundary conditions.

The classical phase space of pure AdS$_3$ gravity is $\left( \faktor{\text{Diff}(\mathbb{S}^1)}{PSL(2;\mathbb{R})}\right)^2$~\cite{cotler2019theory}, and the action~\eqref{E:alekseevShatashvili} is a quantization of that phase space.

We now take the large radius limit of the action above. To do so, we first rescale
\begin{equation}
	t = \frac{u}{L}\,,
\end{equation}
and then make an ansatz for $\phi$ and $\bar{\phi}$ so that the action remains finite in the $L\to\infty$ limit, namely
\begin{equation}
\label{E:AdSansatz}
	\phi(\theta,u) = f\left(\theta-\frac{\alpha(\theta,u)}{L},u\right)\,, \qquad \bar{\phi}(\theta,u) = f\left( \theta+\frac{\alpha(\theta,u)}{L},u\right)\,.
\end{equation}
With these substitutions the action~\eqref{E:alekseevShatashvili} becomes, after suitable integrations by parts,
\begin{align}
\begin{split}
\label{E:AdS3actionLs}
	S &= - \frac{L}{32\pi G} \int d^2 x \, \left(\frac{\phi'' \left(\frac{1}{L}\,\partial_\theta + \partial_u \right) \phi'}{\phi'^2} - \phi' \left(\frac{1}{L}\,\partial_\theta + \partial_u \right) \phi \right.
	\\
	& \qquad \qquad \qquad \qquad \qquad + \left.\frac{\overline{\phi}'' \left(\frac{1}{L}\,\partial_\theta - \partial_u \right) \overline{\phi}'}{\overline{\phi}'^2} - \overline{\phi}' \left(\frac{1}{L}\,\partial_\theta - \partial_u \right) \overline{\phi} \right)
	\\
	& \xrightarrow{L\to\infty} \frac{1}{8\pi G} \int d\theta du\left( \alpha \partial_u \left\{ \tan\left( \frac{f}{2},\theta\right)\right\} + \left\{ \tan\left( \frac{f}{2}\right),\theta\right\}\right)\,.
\end{split}
\end{align}
Upon defining $P = -\frac{1}{8\pi G}\left\{ \tan\left( \frac{f}{2}\right),\theta\right\}$, we recognize this to be just the flat space action~\eqref{E:action1}.

This large radius limit is useful to demonstrate other properties of the flat space theory. For example, the time-dependent $PSL(2;\mathbb{R})\times PSL(2;\mathbb{R})$ gauge redundancy~\eqref{E:AdSgauge} in AdS$_3$ becomes the $ISO(2,1)$ redundancy in flat space~\eqref{E:gauge}. To see this, consider $PSL(2;\mathbb{R})\times PSL(2;\mathbb{R})$ gauge transformations that preserve the large radius form~\eqref{E:AdSansatz} of $\phi$ and $\bar{\phi}$. These are generated by two subgroups. The first are transformations which are $O(1)$, for which $(a,b,c,d)$ must match the barred parameters $(\bar{a},\bar{b},\bar{c},\bar{d})$. These form a $PSL(2;\mathbb{R})$ subgroup and act on $\tan\left( \frac{f}{2}\right)$ by fractional linear transformations
\begin{equation*}
	\tan\left( \frac{f(\theta,u)}{2}\right) \sim \frac{a(u) \tan\left( \frac{f(\theta,u)}{2}\right)+b(u)}{c(u)\tan\left( \frac{f(\theta,u)}{2}\right)+d(u)}\,.
\end{equation*}
The second subgroup is given by group elements given by the identity plus $O(1/L)$ corrections. These take the form
\begin{equation}
	\begin{pmatrix} a & b \\ c & d \end{pmatrix} = \mathbb{1} - \frac{M}{L} + O\left( \frac{1}{L^2}\right) \,, \qquad \begin{pmatrix} \bar{a} & \bar{b} \\ \bar{c} & \bar{d} \end{pmatrix} = \mathbb{1}+\frac{M}{L}+O\left( \frac{1}{L^2}\right)\,,
\end{equation}
with
\begin{equation}
	M = \begin{pmatrix} \delta a(u) & \delta b(u) \\ \delta c(u) & - \delta a(u) \end{pmatrix}\,.
\end{equation}
These form an $\mathbb{R}^3$ subgroup as $L\to\infty$. This transformation leaves $f$ inert, but acts on $\alpha$ as
\begin{equation*}
	\alpha(\theta,u) \sim \alpha(\theta,u) + \frac{\frac{  \delta b(u)+\delta c(u)-2i \delta a(u)}{2}e^{i f(\theta,u)}+\delta b(u)-\delta c(u)+ \frac{  \delta b(u)+\delta c(u)+2i \delta a(u)}{2}e^{-i f(\theta,u)}}{\partial_{\theta}f(\theta,u)}\,,
\end{equation*}
which matches the transformation of $\alpha$ in~\eqref{E:gauge} with the identifications 
\begin{equation*}
	a_1(u) = \frac{\delta b(u)+\delta c(u) - 2 i \delta a(u)}{2}\,, \qquad a_0(u) = \delta c(u) - \delta b(u)\,, \qquad a_{-1}(u) = a_1(u)^*\,.
\end{equation*}
Together these transformations generate the group $ISO(2,1) = PSL(2;\mathbb{R})\ltimes \mathbb{R}^3$.

Furthermore, the invariance of the AdS$_3$ action under $PSL(2;\mathbb{R})\times PSL(2;\mathbb{R})$ gauge transformations implies the invariance of the flat space action $ S_+$ under $ISO(2,1)$ gauge transformations.

The AdS$_3$ action has a traceless stress tensor, with right- and left-moving components
\begin{equation}
	T = -\frac{L}{16\pi G} \left\{ \tan\left( \frac{\phi}{2}\right),\theta\right\}\,, \qquad \bar{T} = - \frac{L}{16\pi G}\left\{ \tan\left( \frac{\bar{\phi}}{2}\right),\theta\right\}\,,
\end{equation}
and the field equations of the model are nothing more than conservation equations,
\begin{equation}
	\partial_+ T = \partial_- \bar{T} = 0\,.
\end{equation}
In the large radius limit a quick computation reveals
\begin{align}
\begin{split}
	\lim_{L\to \infty} \frac{T + \bar{T}}{L}&= - \frac{1}{8\pi G} \left\{ \tan\left( \frac{f}{2}\right),\theta\right\} = P\,,
	\\
	\lim_{L\to\infty} (\bar{T} - T) & = \alpha \partial_{\theta} P + 2P \partial_{\theta} \alpha - \frac{1}{8\pi G} \partial_{\theta}^3 \alpha = J\,,
\end{split}
\end{align}
where $J$ is given in~\eqref{E:J}. The conservation equations $\partial_+T = \partial_- \bar{T}=0$ similarly become the field equations of the flat space action~\eqref{E:flatEoMs} in the large radius limit, 
\begin{align}
\begin{split}
\label{E:flatWard}
	2\lim_{L\to\infty} \frac{(\partial_+ T - \partial_- \bar{T}) }{L^2} &= \partial_u P = 0\,,
	\\
	2\lim_{L\to\infty} \frac{\partial_+ T + \partial_- \bar{T}}{L} & = -\partial_u J + \partial_{\theta} P  = 0\,,
\end{split}
\end{align}

\subsection{Partition function}
\label{sec:partition-function}

In this Subsection, we compute the partition function of pure 3d gravity with vanishing cosmological constant. We have in mind the sum over Euclidean spaces where the boundary is a torus. This partition sum has been computed to one-loop order using a heat kernel in~\cite{Barnich:2015mui}. That result matches the character of the vacuum representation of $\widehat{\text{BMS}}_3$~\cite{Oblak:2015sea}. With the exact Hilbert space and transition amplitudes in hand, we are equipped to compute the result thanks to the boundary description~\eqref{E:action1}. However, we will see that the result has a nowhere-smooth dependence on the rotation parameter, diverging in a complicated way at a dense subset thereof. Our analysis will clear up some extant confusion in the literature, and establish that the partition function is one-loop exact at an irrational rotation parameter.

The partition function of interest is
\begin{equation}
  \label{eq:partition-function}
Z = \text{tr}\!\left(e^{- \beta \widehat{H} + i \omega \widehat{J}}\right)\,.
\end{equation}
The integrality of angular momentum implies that $\omega$ is an angular variable with $\omega \sim \omega + 2\pi$. At $\omega = 0$ we can neglect the angular momentum term and we are simply dealing with
\begin{equation}
\label{E:Z1}
    \text{tr}\!\left( e^{-\beta \widehat{H}}\right) = \int \frac{[dP(\theta)]}{PSL(2;\mathbb{R})}\langle P(\theta)| e^{-\beta \widehat{H}}|P(\theta)\rangle = \delta[0] \int \frac{[dP(\theta)]}{PSL(2;\mathbb{R})}\,e^{\frac{\beta}{8\pi G} \int_0^{2\pi}d\theta \,\left\{ \tan\left( \frac{f(\theta)}{2}\right),\theta\right\}}\,.
\end{equation}
The factor of $\delta[0]$ is familiar in the study of magnetic Carrollian theories~\cite{Cotler:2024xhb}. It should be understood as an IR divergence that arises from the continuous spectrum of energy eigenstates. Stripping it off yields a perfectly finite and sensible partition sum $Z_{\rm finite}$. That finite prefactor closely resembles the Schwarzian path integral~\cite{Stanford:2017thb} familiar from the study of JT gravity, however the measure is somewhat different. It has a saddle-point approximation at large $\frac{\beta}{G}$ where the saddle is $P = -\frac{1}{16\pi G}$ and the measure for small fluctuations around the saddle is $\prod_{|n|>1} dP_n$ for $P_n$ the $n$th Fourier mode of $P(\theta)$. The prefactor is given by
\begin{align}
    \begin{split}
    Z_{\rm finite} &= e^{\frac{\beta}{8G}}\int \prod_{|n|>1} dP_n e^{-\frac{8\pi^2 \beta G |P_n|^2}{n^2-1} + O(\beta G^2 P_n^3)}
    \\
    &= e^{\frac{\beta}{8G}} \prod_{n>1} \frac{n^2-1}{8\pi \beta G}\left( 1 + O\!\left( \frac{G}{\beta}\right)\right)\,,
 \end{split}
\end{align}
Using a Zeta regularization we find that the infinite product evaluates to
\begin{equation}
    Z_{\rm finite} = \pi \left( 8 \pi \beta G\right)^{3/2}e^{\frac{\beta}{8G}}\left( 1+O\!\left( \frac{G}{\beta}\right)\right)\,,
\end{equation}
which, for some normalization of the usual measure for the Schwarzian path integral, coincides with the one-loop approximation to the usual Schwarzian path integral~\cite{Stanford:2017thb}.\footnote{Note that the effective coupling is $\sim G/\beta$, rather than that $\sim \beta J$ familiar from JT gravity and the SYK model. As a result the inverse Laplace transform of $Z_{\rm finite}$ produces a rather different density of states than the usual one associated with the Schwarzian path integral.} We can write this finite answer in terms of a total density of states $\rho(E)$, representing equal contribution from states of all spin at the same energy, via
\begin{equation}
    Z_{\rm finite} = \int_{-\frac{1}{8G}}^{\infty} dE \,\rho(E)\, e^{-\beta E}\,, \qquad \rho(E) = 12 \sqrt{2}\pi G^{3/2}\left( E+\frac{1}{8G}\right)^{-5/2}\,,
\end{equation}
where the integral over $E$ is performed by analytic continuation of the usual integral representation of the Euler Gamma function.

At nonzero rotation parameter, the story is more subtle. Using the transition amplitudes~\eqref{E:Amplitude1} and that $J$ generates a rotation we can write this as
\begin{align}
Z &= \int \frac{[dP(\theta)]}{PSL(2;\mathbb{R})} \,\langle P(\theta)| e^{- \beta \widehat{H} + i \omega \widehat{J}} |P(\theta)\rangle \nonumber \\
  &= \int \frac{[dP(\theta)]}{PSL(2;\mathbb{R})} \, e^{- \beta E[P]}\, \delta[P(\theta) - P(\theta + \omega)]
    \label{eq:partition-1}
\end{align}
for $E[P] =  \int_0^{2\pi}d\theta \, P(\theta)$.  Since $P(\theta)$ is $2\pi$-periodic, observe that for $\frac{\omega}{2\pi} \not \in \mathbb{Q}$, the only solution to $P(\theta) = P(\theta + \omega)$ is $P(\theta) = \text{const}$.  Moreover, the constant is fixed since $P = - \frac{1}{8 \pi G}\{\tan(f/2),\theta\}$ for $f \in \text{Diff}(\mathbb{S}^1)/\text{PSL}(2;\mathbb{R})$; the only possible value is $- \frac{1}{16\pi G}$.  Then we can write
\begin{equation}
\label{E:Zlocalized}
Z = \int \frac{[dP(\theta)]}{PSL(2;\mathbb{R})} \, e^{\frac{\beta}{8 G}} \, \delta[P(\theta) - P(\theta + \omega)]\,.
\end{equation}

The only non-trivial remaining step is to evaluate the integral $\int \frac{[dP(\theta)]}{PSL(2;\mathbb{R})} \, \delta[P(\theta) - P(\theta + \omega)]$, i.e.\ the Jacobian arising from the delta functional.  Notice that the answer must be one-loop exact since the Jacobian is a purely one-loop effect.  We can write the delta functional in its Fourier representation via
\begin{equation}
Z= e^{\frac{\beta}{8G}} \int \frac{[dP(\theta)]  [d\alpha(\theta)]}{ISO(2,1)} \, e^{i \int_0^{2\pi} d\theta \, \alpha(\theta) \, \left(P(\theta) - P(\theta + \omega) \right)}\,.
\end{equation}
Expanding $P(\theta)$ in small fluctuations via $P(\theta) = -\frac{1}{16 \pi G} + \sum_{|n| > 1} \frac{P_n}{\sqrt{2\pi}} \, e^{i n \theta}$ with $|P_n| \ll 1$, we find
\begin{align}
\begin{split}
e^{\frac{\beta}{8G}} \int \frac{[dP]  [d\alpha]}{ISO(2,1)} \, e^{2\pi i \sum_{|n| > 1} \alpha_n \, P_{-n}\,(1 - e^{- i n \omega})} &= e^{\frac{\beta}{8G}} \int \frac{[dP(\theta)]}{PSL(2;\mathbb{R})} \prod_{|n| > 1} \delta(P_n(1 - e^{- in \omega}))\\
&= e^{\frac{\beta}{8 G}} \int \frac{[dP(\theta)]}{PSL(2;\mathbb{R})} \, \prod_{|n| > 1} \delta(P_n) \prod_{n = 2}^\infty \frac{1}{|1-q^n|^2}\,,
\end{split}
\end{align}
where in the last line we have used $q = e^{i \omega}$.  The $\prod_{n = 2}^\infty \frac{1}{|1-q^n|^2}$ term is the Jacobian coming from the delta functional. Near $P = -\frac{1}{16\pi G}$ we have $\frac{[dP(\theta)]}{PSL(2;\mathbb{R})} \approx \prod_{|n|>1} dP_n$ and so integrating over the $P_n$'s, we arrive at the one-loop exact answer
\begin{equation}
\label{E:Zexact1}
Z_{\text{exact}}(\beta, \omega) = e^{\frac{\beta}{8 G}} \prod_{n = 2}^\infty \frac{1}{|1-q^n|^2}\,, \qquad \frac{\omega}{2\pi} \not \in \mathbb{Q}\,.
\end{equation}

So far, we have treated the cases when $\omega =0$ and $\frac{\omega}{2\pi} \not \in \mathbb{Q}$. In the former, all states contribute to the partition sum, while in the latter the only state that contributed was $\ket{P(\theta)=- \frac{1}{16 \pi G}}$. But what if $\frac{\omega}{2\pi}>0$ is rational? We have to be careful in that case; note that if we take $\frac{\omega}{2\pi}$ to be rational in~\eqref{E:Zexact1}, the product has infinitely many poles. The most dramatic case of this is $\omega = 0$, for which all terms in the product are infinite. In that case, the result~\eqref{E:Zexact1} is no longer exact: the divergent product reconstructs the $\delta[0]$ in~\eqref{E:Z1}, and the prefactor $e^{\frac{\beta}{8G}}$ is the tree-level approximation to the residual Schwarzian path integral.

Let us consider rational $\frac{\omega}{2\pi} = \frac{c}{d}$ with $c$ and $d>1$ coprime. The Hilbert space trace localizes to those configurations with $P(\theta + \omega) = P(\theta)$; there is an infinite-dimensional submanifold of such $P(\theta)$. To see this, expand the reparameterization $f(\theta) = \theta + \varepsilon(\theta)$ around the identity, so that
\begin{equation}
	P(\theta) = - \frac{1}{16\pi G} - \frac{\varepsilon'''(\theta) + \varepsilon'(\theta)}{8\pi G} - \frac{1}{16\pi G}(\varepsilon'(\theta)^2 - 3 \varepsilon''(\theta)^2 - 2 \varepsilon'''(\theta)\varepsilon'(\theta)) + O(\varepsilon^3)\,.
\end{equation}
The corrections are multinomials in $\varepsilon$ and its derivatives. Expanding $\varepsilon$ in Fourier modes, $\varepsilon(\theta) = \sum_{|n|>1} \varepsilon_n e^{in\theta}$, those modes with $n = m d $ for $|m|\geq 1$ are periodic with periodicity $\omega$ and we may label the submanifold of configurations with $P(\theta+\omega) = P(\theta)$ by the $\varepsilon_{md}$'s. 

We cannot perform the exact path integral in this setting. However, at small $G$, we can proceed by separating the Fourier modes into the $\varepsilon_{md}$'s, and everything else. Parameterizing $P_n = \frac{i}{8\pi G}n(n^2-1)\varepsilon_n$ as in~\eqref{E:Pmodes} this gives
\begin{align}
\begin{split}
	Z = e^{\frac{\beta}{8G}}\int \prod_{|n|>1} d\alpha_{-n}dP_n(1 + O(G))\exp\!\Big( &  2\pi i \sum_{|n|>1} \alpha_{-n}(P_n+O(GP^2)) (1 - e^{-i n \omega})
	\\
	 & \qquad \qquad - 8\pi^2  \beta G\sum_{|n|>1}\frac{|P_n|^2}{n^2-1}+ O(G^2)\Big)\,.
\end{split}
\end{align}
To leading order in small $G$ the $\alpha_{-n}$ set $P_n$ to vanish for $n \neq m d$. The $\alpha_{-n}$ for $n = m d$ do not appear in the integrand and leading to an IR divergent prefactor $\prod_{n = md >1} \delta(0)^2$. At $n = md$ there is also a residual approximately Gaussian integral over $P_n$. The end result is
\begin{equation}
	Z \approx e^{\frac{\beta}{8G}}\left(  \prod_{n=md>1} \!\delta(0)^2 \,\frac{n^2-1}{8\pi \beta G}\right)\prod_{n>1,\,n\neq m d} \frac{1}{|1-q^n|^2}\,, \qquad \frac{\omega}{2\pi} =\frac{c}{d} \in \mathbb{Q}\,.
\end{equation}
Comparing this result to the case of irrational $\frac{\omega}{2\pi}$ in~\eqref{E:Zexact1}, we see that the poles of the partition sum in~\eqref{E:Zexact1} are broadened into an IR-divergent prefactor $\delta(0)^2$ times a residual, approximately Gaussian, integral over the locus of supermomenta that are periodic modulo $\omega$. 

The IR divergent prefactor is a consequence of a continuous spectrum of states contributing at rational $\frac{\omega}{2\pi}$, namely those states $|P(\theta)\rangle$ with $P(\theta)$ periodic mod $\omega$. From that point of view the factors of $\delta(0)$ arise because of the IR-divergent norm of those states. There is a finite effective statistical system where we strip off those factors, describing the sum over those states. Moreover, whether one uses the complete, divergent answer or its finite effective version, one finds a finite density of states, average energy in the canonical ensemble, and so on.

The partition sum is an extremely complex function of rotation parameter $\omega$. We stress that this complexity is misleading. The underlying quantum mechanics is quite simple: supermomentum eigenstates $|P(\theta)\rangle$ are energy eigenstates with a simple energy $E[P] = \int_0^{2\pi} d\theta\,P(\theta)$. The complicated behavior of $\text{tr}\!\left( e^{-\beta \widehat{H} + i \omega \widehat{J}}\right)$ arises simply because the action of the angular momentum operator $\widehat{J}$ is quite complicated on the supermomentum basis. The quantum mechanics of the boundary gravitons is best described in terms of the $S$-matrix~\eqref{E:Amplitude1} rather than the partition function.

The computation above clears up a subtle point in older computations of the partition sum, in which the rotation parameter $\omega$ was slightly complexified so as to have a very small imaginary part. By~\eqref{E:Zlocalized} the partition sum localizes to configurations $P(\theta)$ that are periodic modulo $\omega$. If $\text{Im}(\omega)$ is nonzero the only such configuration is the constant $P = -\frac{1}{16\pi G}$, in which case we land on the result~\eqref{E:Zexact1} that we found at irrational $\frac{\omega}{2\pi}$. 

\subsection{Summary and results for all of flat space}
\label{S:futureAndPast}

So far, we have discussed the quantum mechanics of boundary gravitons in the future half of three-dimensional asymptotically flat spacetime. The version for the past half of 3d Minkowski space is nearly identical. We already summarized the boundary path integral, action, and boundary conditions in Eqs.~\eqref{E:totalZ},~\eqref{E:totalS}, and~\eqref{E:gluing}.

\begin{figure}
\begin{center}
\includegraphics[scale=.5]{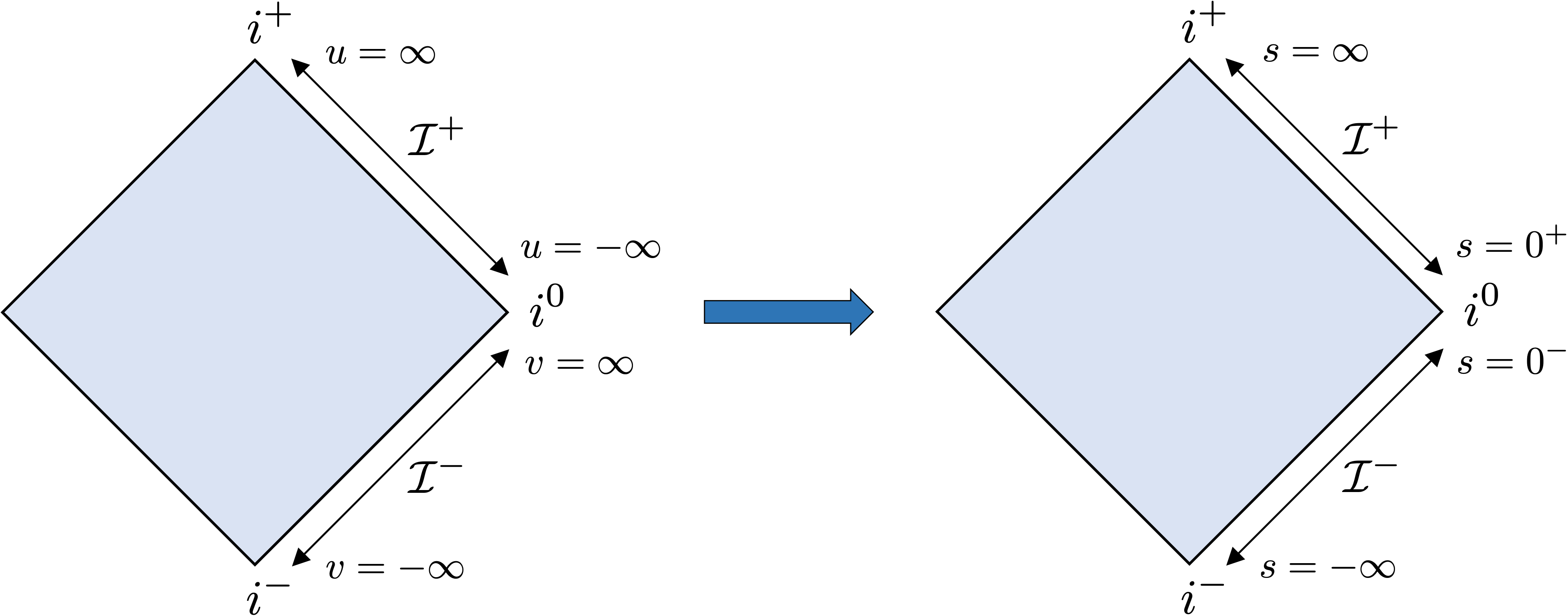}
\end{center}
\caption{A schematic of the change from our original time coordinates $u$ in the future and $v$ in the past to a single global time coordinate $s$. Note that all finite values of $u$ and $v$ are mapped to $s=\infty$ and $s=-\infty$, respectively. \label{fig:coordinate_change} }
\end{figure}

We can combine the two descriptions into a single one with a ``twist'' inserted corresponding to the gluing condition~\eqref{E:gluing}, the three-dimensional version of antipodal matching at spatial infinity. To do so we perform a simple coordinate redefinition. The $u$-coordinate along future null infinity has range $(-\infty, \infty)$, with $-\infty$ corresponding to $i^0$ and $\infty$ corresponding to $i^+$. Let us shift it so that it has range $(0,\infty)$. Similarly, the $v$-coordinate along past null infinity has range $(-\infty, \infty)$, with $-\infty$ corresponding to $i^-$ and $\infty$ corresponding to $i^0$. We shift it so that its range is $(-\infty, 0)$. Since the new ranges of $u$ and $v$ are non-overlapping, we can without confusion encompass both of them by a single coordinate which we call $s$ on $(-\infty, \infty)$.
See Figure~\ref{fig:coordinate_change}.

In terms of this global time coordinate $s$, the antipodal matching conditions are
\begin{equation}
P_+(s \to 0^+, \theta) = P_-(s \to 0^-, \theta + \pi)\,, \qquad \alpha_+(s \to 0^+, \theta) = \alpha_-(s \to 0^-, \theta + \pi)\,.
\end{equation}

We then group the fields together as
\begin{equation}
P(s,\theta) := \begin{cases} P_-(s,\theta)\,, & s < 0 \\
P_+(s,\theta)\,, & s> 0
\end{cases}\,, \qquad \alpha(s,\theta) := \begin{cases} \alpha_-(s,\theta)\,, & s < 0 \\
\alpha_+(s,\theta)\,, & s > 0
\end{cases}\,, 
\end{equation}
so that the antipodal matching corresponds to the ``twist''
\begin{equation}
\label{E:twist}
    P(s\to 0^+,\theta+\pi) = P(s\to 0^-,\theta)\,, \qquad \alpha(s\to 0^+,\theta+\pi) = \alpha(s\to 0^-,\theta)\,.
\end{equation}
These fields are subject to a time-dependent $ISO(2,1)$ identification. The total action is then
\begin{equation}
S_{\text{tot}}[\alpha, f] := \int_{-\infty}^\infty ds \int_0^{2\pi} d\theta \, ( \alpha \,\partial_s P + P)\,.
\end{equation}
This action looks identical to that~\eqref{E:action1} for the future half of flat space, but by virtue of our coordinate redefinitions and the twist it describes all of asymptotically flat spacetime.

The path integral computes transition amplitudes between states in the far past and the far future. We can take them to be supermomentum eigenstates $|P(\theta)\rangle$ with $P\in \faktor{\text{BMS}_3}{ISO(2,1)}$, prepared by Dirichlet boundary conditions on $P$ (really the reparameterization $f$). Because these states are energy eigenstates with a non-trivial energy function $E[P]$ it is convenient to cut off time in the past and future so that we work with $s \in [-T/2, T/2]$, with evolution from $-T/2$ to $T/2$ generated by the evolution operator $\widehat{\mathcal{U}}(T/2,-T/2)$. This allows us to focus on a neighborhood of spatial infinity.  Using the transition amplitudes~\eqref{E:Amplitude1} and the antipodal matching condition~\eqref{E:twist} we then have
\begin{equation}
    \langle P_1(\theta)|\,\widehat{\mathcal{U}}(T/2,-T/2)|P_2(\theta)\rangle = \delta[P_1(\theta)-P_2(\theta+\pi)]e^{-i E[P_2]T}\,, \qquad E[P] = \int_0^{2\pi} d\theta\,P(\theta)\,.
\end{equation}
In this form the conservation of supermomentum (modulo antipodal matching) is manifest. This selection rule is the analogue of the soft theorem of the next Section in the absence of matter.

Note that here we have found it prudent to quantize pure gravity in the $u$-direction, i.e.~where $u$ is a time coordinate.  When we couple gravity to massless matter in Section~\ref{sec:adding-matter}, it will be more convenient to quantize in the $r$-direction.  Then we will define our initial states on the surface where $r \to \infty$ and $-\infty < s < 0$, and our final states on the surface where $r \to \infty$ and $0 < s < \infty$.  For more details see Section~\ref{subsubsec:Hilbert1}.

In this Section we have quantized pure gravity on spacetimes that are diffeomorphic to Minkowski space. A generalization of these methods presumably gives a boundary description of Minkowski space with a conical deficit, and of flat space cosmologies, building off of~\cite{Barnich:2017jgw,Merbis:2019wgk, Prohazka:2017equ,Compere:2017knf}.  Perhaps it is also possible to obtain boundary graviton descriptions when we relax the boundary conditions to allow for more general falloffs than those we considered~\eqref{E:asymptoticallyflat1}.

\section{Adding matter}
\label{sec:adding-matter}

In this Section, our primary goal is to derive an analogue of the infrared triangle~\cite{Strominger:2017zoo} for three-dimensional gravity coupled to massless matter. We indeed find soft theorems, asymptotic symmetries, and memory effects, organized into a triangle as depicted in Figure~\ref{fig:3d-triangle}. The vertices of the triangle can be thought of as manifestations of sourced Ward identities at infinity.  As such it can be useful to regard the three-dimensional IR triangle as an infrared trivalent vertex.

We proceed with a brief discussion of the Ward identities, followed by their connection to asymptotic symmetries, soft theorems, and memory effects. As such, three-dimensional gravity evidently has more in common with four-dimensional gravity than previously realized.

\subsection{Ward identities}
\label{subsec:Ward1}

A particularly transparent approach to deriving the Ward identities is to consider the far-field limit of the constraints of Einstein gravity coupled to massless matter. We do so using the Einstein equations in combination with appropriate fall-off conditions of the metric near infinity. Our starting point is the Einstein-Hilbert action coupled to a massless scalar field $\Phi$, namely
    \begin{equation}
        S=\int d^3 x \sqrt{-g}\left(\frac{1}{16 \pi G}\, R-\frac{1}{2}(\partial\Phi)^2\right)\,.
    \end{equation}
Now let us introduce Bondi coordinates $0<r<\infty$, $-\infty<u<\infty$, $0\leq\theta\leq2\pi$ and work in a gauge where the metric takes the form 
    \begin{equation}
    \label{eq:ScalarBCs}
        d s^2 = e^{2\beta}\frac{V}{r}d u^2-2e^{2\beta}d ud r+r^2(d\theta-Ud u)^2\,,
    \end{equation}
where the fields $\beta$, $U$ and $V$ fall off as $\beta=\mathcal{O}(r^{-1})$, $U=\mathcal{O}(r^{-2})$, $V=\mathcal{O}(r)$~\cite{Barnich:2013yka}. We parameterize these falloffs as
\begin{align}
\begin{split}
\label{E:asym2}
    \beta(u,r,\theta) &= \frac{L_{-1}(u,\theta)}{r} + \frac{L_{-2}(u,\theta)}{r^2} + \cdots\,,
    \\
    V(u,r,\theta) &= M_1(u,\theta)\,r + M_0(u,\theta) + \cdots \,,
    \\
    U(u,r,\theta) &= - \frac{N_{-2}(u,\theta)}{r^2} - \frac{N_{-3}(u,\theta)}{r^3} - \cdots\,.
\end{split}
\end{align}
Comparing to the form of~\eqref{E:flatsolutions}, we identify
\begin{align}
M_1(u,\theta) := 16 \pi G\,P(u,\theta)\,,\quad N_{-2}(u,\theta) := 8 \pi G \, J(u,\theta)\,.
\end{align}

As for the falloff for the stress tensor of a massless field, we parameterize it near null infinity by
\begin{align}
\begin{split}
T_{u u}(u,r,\theta) &= \frac{T_{uu}^{(-1)}(u,\theta)}{r} + \frac{T_{uu}^{(-2)}(u,\theta)}{r^2} + \cdots\,,
\\
T_{r u}(u,r,\theta)& = \frac{T_{ru}^{(-3)}(u,\theta)}{r^3} + \frac{T_{ru}^{(-4)}(u,\theta)}{r^4}  + \cdots\,, 
\\ 
T_{r r}(u,r,\theta) &= \frac{T_{rr}^{(-3)}(u,\theta)}{r^3} + \frac{T_{rr}^{(-4)}(u,\theta)}{r^4}  + \cdots\,,
\\
T_{r \theta}(u,r,\theta) &= \frac{T_{r\theta}^{(-2)}(u,\theta)}{r^2} + \frac{T_{r\theta}^{(-3)}(u,\theta)}{r^3}  + \cdots\,, 
\\
T_{\theta \theta}(u,r, \theta) &= T_{\theta\theta}^{(0)}(u,\theta) + \frac{T_{\theta\theta}^{(-1)}(u,\theta)}{r} + \cdots \,,
\\
T_{u \theta}(u,r,\theta) &= \frac{T_{u\theta}^{(-1)}(u,\theta)}{r} + \frac{T_{u\theta}^{(-2)}(u,\theta)}{r^2} + \cdots\,.
\end{split}
\end{align}
Then the constraint components of the Einstein equations at large $r$ become evolution equations in retarded time $u$ of $P$ and $J$:
\begin{equation}
\label{eq:boxedEOM}
\boxed{
\begin{aligned}
            \partial_u P &= -T_{uu}^{(-1)}\,,
            \\
            \partial_u J &= \partial_\theta \!\left(P - \frac{1}{2}\, T_{\theta \theta}^{(0)}\right) - T_{u \theta}^{(-1)}\,.
\end{aligned}
}
\end{equation}
In the absence of matter \eqref{eq:boxedEOM} leads to the conservation equations for supermomentum and super-angular momentum.  In the presence of matter energy-momentum, they are modified as above, and indeed, in the quantum theory the relations~\eqref{eq:boxedEOM} are promoted to operator equations and become our desired Ward identities.

Now we specialize to the stress tensor of a minimally coupled free scalar,
\begin{equation}\label{eq:ScalarStressEnergyTensor}
        T_{\mu\nu}=\partial_\mu\Phi\partial_\nu\Phi-\frac{1}{2}g_{\mu\nu}(\partial\Phi)^2\,.
    \end{equation}
The classical EOMs are given by the Einstein equations plus the Klein-Gordon equation for the scalar field
    \begin{equation}
        R_{\mu\nu}-\frac{1}{2}g_{\mu\nu}R=8 \pi G\, T_{\mu\nu}\,,\qquad \square \Phi=0\,.
    \end{equation}
To solve these equations asymptotically we impose the appropriate falloff conditions on the scalar field,
\begin{align}
\Phi=\Phi_{-1/2}(u,\theta) \,r^{-1/2}+\Phi_{-3/2}(u,\theta) \,r^{-3/2}+ \cdots\,.    
\end{align}
For ease of notation, we will define $\Phi_0 := \Phi_{-1/2}$ and $\Phi_1 := \Phi_{-3/2}$, $\Phi_2 := \Phi_{-5/2}$, and so on.  With the above fall-off behavior, the asymptotic form of the stress-energy tensor \eqref{eq:ScalarStressEnergyTensor} is
\begin{align}
\label{eq:leadingSET}
\begin{split}\
            T_{rr} & = \frac{\Phi_0^2}{4r^3} + \mathcal{O}\!\left(r^{-\frac{7}{2}}\right)\,,
            \hspace{.64in}
            T_{ru} = \frac{-4 \pi G\,P\, \Phi_0^2 + (\partial_\theta\Phi_0)^2}{2r^3} + \mathcal{O}\!\left(r^{-4}\right)\,,
            \\
            T_{r\theta} &= \frac{-\Phi_0\partial_\theta\Phi_0}{2r^2} + \mathcal{O}\!\left(r^{-3}\right)\,,
            \qquad
            T_{uu} = \frac{(\partial_u\Phi_0)^2}{r} + \mathcal{O}\!\left(r^{-2}\right)\,,
            \\
            T_{u\theta} &= \frac{\partial_u\Phi_0\partial_\theta\Phi_0}{r} + \mathcal{O}\!\left(r^{-2}\right)\,, 
            \qquad \hspace{-.01in}
            T_{\theta\theta} = -\frac{1}{2}\Phi_0\partial_u\Phi_0 + \mathcal{O}\!\left(r^{-1}\right). 
\end{split}
\end{align}
Solving the Einstein equations asymptotically fixes $\beta$, $U$, and $V$ for large $r$ as
\begin{align}
\begin{split}
            \beta &=  -\frac{\pi G}{r}\Phi_0^2-\frac{3 \pi G}{r^2}\Phi_0\Phi_1+\cdots\,,
            \\
            U &= -\frac{8 \pi G\,J}{r^2} + \frac{2 \pi G}{3r^3}(5\Phi_0\partial_\theta\Phi_1-3\partial_\theta\Phi_0\Phi_1+16 \pi G J\,\Phi_0^2)+\cdots\,,
            \\
            V &= 16\pi G\, P\,r + 4 \pi G (\partial_\theta^2\Phi_0\Phi_0-(\partial_\theta\Phi_0)^2)+ \cdots\,,
\end{split}
\end{align}
where $P$ and $J$ satisfy~\eqref{eq:boxedEOM}. 

\subsection{Asymptotic symmetries}
\label{subsec:asympt1}

The Ward identities in~\eqref{eq:boxedEOM} are a consequence of asymptotic symmetries that persist even in the presence of ingoing/outgoing matter. In this Subsection, we demonstrate that this is the case by (1) deriving these asymptotic symmetries and (2) obtaining their corresponding asymptotic charges using the Iyer-Wald prescription.

To determine the asymptotic charges, one first has to determine the asymptotic Killing vectors that leave the boundary conditions~\eqref{eq:ScalarBCs} and~\eqref{E:asym2} invariant (up to variations of the fields $M$, $N$ and $\Phi_0$). It is a straightforward computation to show that, even in the presence of matter parameterized by $\Phi_0$, these asymptotic Killing vectors are given by 
    \begin{equation}\label{ed:ScalarAKV}
        \xi=\left(\xi^r_1 r + \xi^r_0 + \frac{\xi^r_{-1}}{r} + O(r^{-2})\right)\partial_r +( \xi^u_0 + O(r^{-1}))\,\partial_u + \left(\xi^\theta_0 + \frac{\xi^\theta_{-1}}{r} + \frac{\xi^\theta_{-2}}{r^2} + O(r^{-3})\right)\partial_\theta\,,
    \end{equation}
where
\begin{align}
\begin{split}
        \xi^r_1 &= -Y'\,,\hspace{1.08in} \xi^r_0 = T'' + u Y'''\,,
        \\
        \xi^r_{-1} &= -8 \pi G  \left(J + \frac{1}{4}\Phi_0\partial_\theta\Phi_0\right)(T'+uY'') - G \pi \Phi_0^2(T''+uY''')\,,
        \\
         \xi^u_0 &= T + uY',\hspace{.85in}
            \xi^\theta_0 = Y\,,
        \\
        \xi^\theta_{-1} &= -T' - uY''\,,\hspace{.56in}
        \xi^\theta_{-2} = G\pi\,\Phi_0^2(T'+uY'')\,,
\end{split}
\end{align}
where $T=T(\theta)$ and $Y=Y(\theta)$ and a prime denotes a derivative with respect to $\theta$. Note the dependence on the matter configuration in the subleading terms for $\xi^r$ and $\xi^{\theta}$. The Lie bracket of the asymptotic Killing vectors obeys the BMS$_3$ algebra (without a central extension)
    \begin{align}
    \begin{split}\label{eq:ScalarAKVAlgebra}
        [\xi[T_1,Y_1],\xi[T_2,Y_2]]&=\xi[T_{12},Y_{12}]\,,
        \\
        T_{12} &= T_1Y_2'-Y_2T_1'-T_2Y_1'+Y_1T_2'\,,
        \\
        Y_{12} &= Y_1Y_2'-Y_2Y_1'\,.
    \end{split}
    \end{align}
Acting with these Killing vectors on a metric that satisfies the boundary conditions~\eqref{eq:ScalarBCs} and~\eqref{E:asym2} we find that the leading behavior of the metric and scalar field vary as
\begin{align}
\begin{split}
   \label{eq:InfinitesimalTransformationsMNScalarField}
        \delta_{\xi} P &= 2PY'+YP'-\frac{Y'''}{8\pi G}+(T+uY')\partial_u P\,,
        \\
        \delta_{\xi} J &= 2JY'+YJ'-\frac{T'''+uY''''}{8\pi G} + (T+uY')\partial_u J + 2(T'+uY'')P
        \\
        &  \quad+\frac{1}{4}(T'+uY'')\Phi_0\partial_u\Phi_0-\frac{1}{8}Y''\Phi^2_0\,,
        \\
        \delta_{\xi}\Phi_0 &= \frac{1}{2}\Phi_0Y'+Y\partial_\theta\Phi_0+(T+uY')\partial_u\Phi_0\,.
\end{split}
\end{align}

Having determined the asymptotic Killing vectors, the next step is to determine the corresponding asymptotic boundary charges. We do so by using the Iyer–Wald prescription; see \cite{Compere:2018aar,Ruzziconi:2019pzd,Frodden:2019ylc} for pedagogical introductions.

In the covariant phase space approach, the infinitesimal surface charge $\slashed{\delta}H_\xi[\mathcal{X}, \delta\mathcal{X}]$ associated with a diffeomorphism $\xi$ is given by
    \begin{equation}\label{eq:InfintesimalSurfaceCharge}
        \slashed{\delta}H_\xi[\mathcal{Y}, \delta\mathcal{X}] = \int_{\partial\Sigma}\,k_\xi[\mathcal{X}, \delta\mathcal{X}] = -\int_{\partial\Sigma}\,(\delta Q_\xi[\mathcal{X}] + i_\xi \theta[\mathcal{X},\delta\mathcal{X}]).
    \end{equation}
Here, $Q_\xi$ is the Noether surface charge, $\theta$ is the symplectic potential, and $\mathcal{X} \equiv \{M, N, \Phi_0\}$ are the coordinates on the solution space. $Q_\xi$ and $\theta$ are to be understood as $n - 2$ forms. 

The infinitesimal surface charge \eqref{eq:InfintesimalSurfaceCharge} is said to be integrable if it is $\delta$-exact, i.e. if
$\slashed{\delta}H_\xi = \delta H_\xi$. The symbol $\slashed{\delta}$ in \eqref{eq:InfintesimalSurfaceCharge} emphasizes that the infinitesimal surface charge is not necessarily integrable. For the Einstein–Hilbert action (without additional matter sources), the Iyer–Wald prescription yields the following (possibly non-integrable) infinitesimal charge
    \begin{equation}
         \slashed{\delta}H_\xi^{EH} = \frac{1}{8\pi G} \int_{\partial\Sigma}(d^{n-2}x)_{\mu\nu}\, \sqrt{-g}(\xi^\mu\nabla_\sigma h^{\nu\sigma} - \xi^\mu\nabla^\nu h + \xi_\sigma\nabla^\nu h^{\mu\sigma} + \frac{1}{2} h \nabla^\nu\xi^\mu  -h^{\sigma\nu}\nabla_\sigma\xi^\mu)\,,
    \end{equation}
where we set $h_{\mu\nu}=\delta_{\xi} g_{\mu\nu}$ to avoid confusion. 
Note that the Barnich–Brandt and Iyer–Wald prescriptions coincide in Bondi gauge \cite{Ruzziconi:2019pzd}. In the case where the Einstein-Hilbert action is supplemented with additional scalar matter degrees of freedom, there is an additional contribution to the charges coming from the matter with symplectic potential
    \begin{equation}
    \label{eq:symppotential}
        \theta[\Phi_0,\delta\Phi_0] = -(d^2x)_\mu\,\sqrt{-g}\,\delta\Phi_0\partial^\mu\Phi_0\,.
    \end{equation}
The Noether surface charge $Q_\xi$ vanishes, so one finds an additional non-integrable contribution
    \begin{equation}
        \slashed{\delta}H_\xi^{S} = -\int_{\partial\Sigma}\,i_\xi \theta[\Phi_0,\delta\Phi_0]\,.
    \end{equation}
See \cite{Frodden:2019ylc,Barnich:2002pi} for explicit expressions of the relevant charge in this context.
Thus, with our choice of boundary conditions, we find for the full charge associated to the asymptotic Killing vector \eqref{ed:ScalarAKV}
    \begin{equation}
        \slashed{\delta}H_\xi = \slashed{\delta}H_T + \slashed{\delta}H_Y,
    \end{equation}
where
\begin{align}
\begin{split}
    \slashed{\delta}H_T &=\int_0^{2\pi} d\theta\, T(\delta P+\partial_u\Phi_0\delta\Phi_0)\,,
    \\
    \slashed{\delta}H_Y &= \int_0^{2\pi} d\theta \left(Y(\delta J-u\delta P')-\frac{1}{8} Y'\left(\delta(\Phi_0^2)-8u\partial_u\Phi_0\delta\Phi_0\right)\right)\,.
\end{split}
\end{align}
It is not surprising that these charges are non-integrable as this problem is one of leaky boundary conditions for the scalar field. To deal with this non-integrability we follow the construction presented in \cite{Barnich:2011mi}. Let $s=(T,Y)$ and split our charges into an integrable $Q_s$ and  a non-integrable piece $\theta_s[\mathcal{X},\delta\mathcal{X}]$ as
    \begin{equation}
    \label{eq:ScalarFieldChargeSplit}
        \slashed{\delta}H_\xi[\mathcal{X},\delta\mathcal{X}] = \delta(Q_s[\mathcal{X}]) + \theta_s[\mathcal{X},\delta\mathcal{X}]\,,
    \end{equation}
where
\begin{align}
\begin{split}
\label{eq:ScalarFieldCharges}
            Q_T[\mathcal{X}] &= \int_0^{2\pi} d\theta\, T P\,, 
            \hspace{2in} \theta_T[\mathcal{X},\delta\mathcal{X}] = \int_0^{2\pi} d\theta\, T\partial_u\Phi_0\delta\Phi_0\,,
            \\
            Q_Y[\mathcal{X}] &=\int_0^{2\pi} d\theta \left(Y(J-u P')- Y'\frac{\Phi_0^2}{8}\right)\,,
            \hspace{.65in}
            \theta_Y[\mathcal{X},\delta\mathcal{X}] = \int_0^{2\pi} d\theta\,  uY'\partial_u\Phi_0\delta\Phi_0\,.
\end{split}
\end{align}
It should be noted that this split into integrable and non-integrable parts is ambiguous, as one can redefine $Q'_s = Q_s - \Tilde{Q}_s$ together with $\theta_s' = \theta_s + \delta\Tilde{Q}_s$ for some $\Tilde{Q}_s[\mathcal{X}]$ which would still satisfy \eqref{eq:ScalarFieldChargeSplit}. However, as we show in a moment the choice of split that we make here leads to a particularly simple asymptotic symmetry algebra.

These charges do not generate the algebra of vector fields \eqref{eq:ScalarAKVAlgebra} under the usual Poisson bracket. However, when defining instead the Barnich-Troessaert bracket \cite{Barnich:2011mi}
    \begin{equation}
    \label{eq:BTbracket}
        \{Q_{\xi_1},Q_{\xi_2}\}^* = \delta_{\xi_2}Q_{\xi_1} + \theta_{\xi_2}[\delta_{\xi_1}\mathcal{X},\mathcal{X}]\,,
    \end{equation}
one finds on-shell
    \begin{equation}
        \{Q_{\xi_1},Q_{\xi_2}\}^* = Q_{[\xi_1,\xi_2]}+K_{\xi_1,\xi_2}[\mathcal{X}]\,,
    \end{equation}
where $K_{\xi_1,\xi_2}[\mathcal{X}]$ is a (possibly field-dependent) co-cycle. Computing these brackets for \eqref{eq:ScalarFieldCharges} one obtains first
    \begin{align}
    \begin{split}
         \{Q_{T_1},Q_{T_2}\}^* &=\int_0^{2\pi} d\theta\,(T_1 \delta_{T_2}P + T_2\partial_u\Phi_0 \delta_{T_1}\Phi_0)
         \\
         &=\int_0^{2\pi} d\theta\,T_1T_2(\partial_u P+(\partial_u\Phi_0)^2)=0\,,
    \end{split}
    \end{align}
where in the last line we have gone on-shell using \eqref{eq:boxedEOM}. For the next bracket, we have
    \begin{align}
    \begin{split}
         \{Q_{T},Q_{Y}\}^* &= \int_0^{2\pi} d\theta\,( T \delta_{Y}P +  u Y'\partial_u\Phi_0 \delta_{T}\Phi_0)
         \\
         &=\int_0^{2\pi} d\theta\,(2PTY'+P'TY-\frac{1}{8\pi G}TY'''+u TY'(\partial_u P+(\partial_u\Phi_0)^2))
         \\
         &=\int_0^{2\pi} d\theta\,(P(TY'-T'Y)-\frac{1}{8\pi G}TY''')=Q_{[T,Y]}-\frac{1}{8\pi G}\int_0^{2\pi} d\theta\,TY'''\,,
    \end{split}
    \end{align}
where, again, in the last line we have gone on-shell. The final bracket requires a little bit more effort to compute, but after imposing the equations of motion and integration by parts one obtains
    \begin{align}
    \begin{split}
         \{Q_{Y_1},Q_{Y_2}\}^*= &\int_0^{2\pi} d\theta \left(Y_1(\delta_{Y_2}J-u \delta_{Y_2}P')- Y'_1\frac{\Phi_0 \delta_{Y_2}\Phi_0}{4}+u Y'_2 \partial_u\Phi_0 \delta_{Y_1}\Phi_0\right)
         \\
         &=\int_0^{2\pi} d\theta \left(Y_{12}(J-u P')-Y'_{12} \frac{\Phi^2_0}{8}\right)
         =Q_{[Y_1,Y_2]}\,,
    \end{split}
    \end{align}
with $Y_{12}=Y_1Y'_2-Y'_1Y_2$.

In summary, we have the on-shell relations
\begin{equation}
    \boxed{
        \begin{aligned}
            \{Q_{T_1}, Q_{T_2}\}^* &= 0\,, 
            \\
            \{Q_{T}, Q_{Y}\}^* &= Q_{[T,Y]}-\frac{1}{8\pi G}\int_0^{2\pi}d\theta\,TY'''\,,
            \\
            \{Q_{Y_1}, Q_{Y_2}\}^* &= Q_{[Y_1, Y_2]}\,,
        \end{aligned}
    }
    \label{eq:boxedPoissonBrackets}
\end{equation}
which are precisely the brackets of the centrally extended BMS$_3$ algebra, with the same central charge as without matter.\footnote{While the bracket \eqref{eq:BTbracket} depends on the splitting, the Koszul bracket has been put forward as an alternative that is independent of this choice \cite{Barnich:toappkb}; see \cite{Bosma:2023sxn} for the explicit expression. It can be checked that the algebra does not change when using this split-independent bracket.} This is not guaranteed since in general one expects field-dependent cocycles to arise, as is the case in four-dimensional gravity \cite{Barnich:2010eb}; see also \cite{Barnich:2015jua, Bosma:2023sxn} for examples in the context of 3d Einstein--Maxwell theory.\footnote{In \cite{Bosma:2023sxn}, two different sets of boundary conditions are considered. In the first set, the fall-offs in the metric include logarithmic terms. In this case, a field-dependent co-cycle arises in the $\textrm{BMS}_3$ algebra. On the other hand, if the fall-offs in the metric are identical to  to the ones chosen in \eqref{E:asym2}, this co-cycle is absent.}

The conservation of the charges with respect to the retarded time $u$ can be checked using
    \begin{equation}
        \frac{d Q}{d u} = \frac{\partial Q}{\partial u} + \delta_{T=1}Q\,.
    \end{equation}
With this, we find for the supermomentum charge
    \begin{equation}
    \label{eq:QTconservation}
        \frac{d Q_T}{d u} = \int_0^{2\pi} d\theta\, T\partial_u P = -\int_0^{2\pi} d\theta\, T (\partial_u\Phi_0)^2 = -\int_0^{2\pi} d\theta\, T\,T_{uu}^{(-1)}\, ,
    \end{equation}
while for the superrotation charge, we get
    \begin{align}
    \label{eq:QYconservation}
    \begin{split}
        \frac{d Q_Y}{d u} &= \int_0^{2\pi} d\theta\,\left(- Y P'+Y(\partial_uJ-u \partial_u P')-Y'\frac{\Phi_0\partial_u\Phi_0}{4}\right)
        \\
        &=-\int_0^{2\pi} d\theta \, Y\left(\partial_\theta T^{(0)}_{\theta\theta}+T^{(-1)}_{u\theta}-u  \partial_\theta T_{uu}\right)\,. 
    \end{split}
    \end{align}
These flux equations are equivalent to the Ward identities~\eqref{eq:boxedEOM} upon identifying $Q_T$ and $Q_Y$ as in~\eqref{eq:ScalarFieldCharges}.

Integrating both sides of the flux equations one finds
\begin{equation}
\begin{split}
    \label{eq:fluxintegrals}
    \lim_{u\to\infty}Q_T-\lim_{u\to-\infty}Q_T&=-\int^{\infty}_{-\infty}du\int_0^{2\pi} d\theta\, T\,T_{uu}^{(-1)}\,,
    \\
    \lim_{u\to\infty}Q_Y-\lim_{u\to-\infty}Q_Y&=-\int^{\infty}_{-\infty}du\int_0^{2\pi} d\theta \, Y\left(\partial_\theta T^{(0)}_{\theta\theta}+T^{(-1)}_{u\theta}-u  \partial_\theta T_{uu}\right)\,.
    \end{split}
\end{equation}
In analogy to the four-dimensional case, one can denote the left and right-hand sides of these equations as soft and hard contributions, respectively. The latter acts on the matter sectors whereas the former acts only on the soft vacuum sector, assuming a compactly supported matter source. 

One can check explicitly that the hard contributions generate the same charge algebra. From the symplectic potential \eqref{eq:symppotential} one constructs the symplectic structure
\begin{equation}
    \Omega(\delta_1\Phi_0,\delta_2\Phi_0)=\int_{-\infty}^{\infty}du \int_0^{2\pi} d\theta \left(\delta_1\Phi \partial_u\delta_2\Phi-\delta_2\Phi \partial_u\delta_1\Phi\right)\,,
\end{equation}
which yields the fundamental Poisson brackets
\begin{equation}
    \{\partial_u\Phi_0(u,\theta),\Phi_0(u',\theta')\}=\frac{1}{2}\delta(u-u')\delta(\theta-\theta')\,.
\end{equation}
Using the explicit expression for the leading components of the stress
tensor \eqref{eq:leadingSET}, one finds the Poisson brackets of the
asymptotic stress tensor with itself
\begin{align}
     \{T^{(-1)}_{uu}(u,\theta),T^{(-1)}_{uu}(u',\theta')\}&=\left(\partial_{u'}-\partial_{u}\right) \left(T^{(-1)}_{uu}\delta(u-u')\delta(\theta-\theta')\right)\,,
     \nonumber
     \\
     \{T^{(-1)}_{uu},T^{(-1)}_{\theta\theta}\}&=T^{(-1)}_{\theta\theta}\partial_{u'}\delta(u-u')\delta(\theta-\theta')\,,
     \nonumber
     \\
     \{T^{(-1)}_{uu},T^{(0)}_{u\theta}\}&=\left(T^{(0)}_{u\theta}\partial_{u'}-\partial_\theta T^{(-1)}_{uu}+T^{(-1)}_{uu}\partial_{\theta'}\right)\delta(u-u')\delta(\theta-\theta')\,,
     \nonumber
     \\
     \{T^{(-1)}_{\theta\theta},T^{(0)}_{\theta\theta}\}&=\frac{1}{4}\partial_{u}\partial_{u'}\!\left(\Phi_0(u,\theta)\Phi_0(u',\theta)\textrm{sgn}(u-u')\right)\delta(\theta-\theta')\,,
     \\
     \{T^{(-1)}_{u\theta},T^{(0)}_{u\theta}\}&=\frac{1}{2}\left(T^{(-1)}_{u\theta}\partial_{\theta'}-T^{(-1)}_{u\theta}(u',\theta')\partial_{\theta'}+T^{(-1)}_{u\theta}-2 \partial_\theta T^{(0)}_{\theta\theta}\right)\delta(u-u')\delta(\theta-\theta')\nonumber\\
&\quad +\frac{1}{4}\partial_u\Phi_0\partial_{u'}\Phi_0\,\textrm{sgn}(u-u')\partial_\theta\partial_{\theta'}\delta(\theta-\theta')+\partial_{u'}\!\left((\partial_\theta\Phi_0)^2\delta(u-u')\delta(\theta-\theta')\right)
     \nonumber
     \\
     \{T^{(-1)}_{u\theta},T^{(0)}_{\theta\theta}\}&=\frac{1}{2}T^{(-2)}_{r\theta}\partial_{u'}\delta(u-u')\delta(\theta-\theta') \nonumber\\
     &\quad -\frac{1}{8}\partial_{u'}\left(\partial_u\Phi_0(u,\theta)\Phi_0(u',\theta')\textrm{sgn}(u-u')\partial_\theta\delta(\theta-\theta')\right),
     \nonumber
\end{align}
where quantities on the right-hand side depend on $u,\theta$ unless otherwise specified.  Using these expressions one confirms that the light-ray integrals of the stress tensor components in \eqref{eq:fluxintegrals} generate the $\widehat{\text{BMS}}_3$ algebra \eqref{eq:boxedPoissonBrackets}.

\subsection{Soft theorems}
\label{subsec:soft1}
\subsubsection{Hilbert space at infinity}
\label{subsubsec:Hilbert1}

To derive soft theorems we need to first discuss a basis of states at future and past null infinity which describes a joint configuration of the gravitational field and a massless matter field. We will work with the global time coordinate $s$ defined in Subsection~\ref{S:futureAndPast}. First, consider a fiducial Hilbert space for configurations along past null infinity, $\mathcal{H}_{\text{past}}^{\text{fiducial}} \simeq \text{span}\{|P(s,\theta), \Phi(s,\theta)\rangle\}$ where $P(s,\theta), \Phi(s,\theta)$ are defined for $-\infty < s \leq 0$.  We take $\Phi(s,\theta)$ to be compactly supported in $s$ so that we do not put massless sources at past timelike infinity.  It is useful to view $\mathcal{H}_{\text{past}}^{\text{fiducial}}$ as
\begin{align}
\mathcal{H}_{\text{past}}^{\text{fiducial}} \simeq \bigotimes_{\substack{s \in (-\infty,0] \\ \theta \in [0,2\pi)}} \mathcal{H}_{s,\theta}\,,
\end{align}
i.e.~as a tensor product of local Hilbert spaces at each $s \in (-\infty, 0]$ and $\theta$. Here $s$ and $\theta$ play the role of spatial coordinates at infinity which label each tensor factor. The above tensor product factorization will be absent as soon as we impose the Ward identities and pass to the constrained space of physical states. 

To proceed we impose the operator Ward identities~\eqref{eq:boxedEOM},  $\widehat{\mathcal{C}}_1 = \partial_s \widehat{P} +\widehat{T}_{ss}^{(-1)} = 0$ and $\widehat{\mathcal{C}}_2 = \partial_s \widehat{J} - \partial_\theta(\widehat{P} - \frac{1}{2}\,\widehat{T}_{\theta\theta}^{(0)}) + \widehat{T}_{s\theta}^{(-1)} = 0$ as constraints annihilating physical states.  This results in the space of physical states $\mathcal{H}_{\text{past}} \subset \mathcal{H}_{\text{past}}^{\text{fiducial}}$
\begin{align}
\mathcal{H}_{\text{past}} \simeq \big\{|\Psi\rangle \in \mathcal{H}_{\text{past}}^{\text{fiducial}} \,:\, \widehat{\mathcal{C}}_1 |\Psi\rangle = \widehat{\mathcal{C}}_2 |\Psi\rangle = 0 \big\}\,.
\end{align}
In the absence of matter the second constraint is redundant, and the first constraint reproduces the fact that $P(s,\theta)$ is constant along null infinity. On states in the fiducial Hilbert space, $\widehat{J}$ acts as the identity on the matter part of the state and in the same way as in pure gravity on the supermomentum part of the state. In the perturbative quantization this action was given in the discussion starting with~\eqref{E:Jn}.  We also define a `future' Hilbert space $\mathcal{H}_{\text{future}}$ in the way we defined $\mathcal{H}_{\text{past}}$.

In total, the full Hilbert space of null infinity is given by
\begin{align}
\mathcal{H}_{\text{tot}} \simeq \mathcal{H}_{\text{future}} \otimes \mathcal{H}_{\text{past}}\,,
\end{align}
with a corresponding algebra of operators denoted by
\begin{align}
\mathcal{A}_{\text{tot}} \simeq \mathcal{A}_{\text{future}} \otimes \mathcal{A}_{\text{past}}\,,
\end{align}
where $\mathcal{A}_{\text{future}} \simeq \mathcal{A}_{\text{past}} \simeq \mathcal{A}$.
There is a canonical operator-ordering map (see e.g.~\cite{Feynman:1951gn, johnson2015feynman} for a more general discussion of Feynman's operator calculus)
\begin{align}
\label{E:Kop1}
\widehat{K} : \mathcal{A}_{\text{future}} \otimes \mathcal{A}_{\text{past}} \to \mathcal{A}
\end{align}
given by
\begin{align}
\label{E:Kaction1}
\widehat{K}(a_f \otimes a_p) = a_f \cdot \widehat{S} \cdot a_p\,,
\end{align}
where $\widehat{S}$ is the $S$-matrix, which includes an antipodal identification of $\theta \to \theta+\pi$.  The right-hand side of~\eqref{E:Kaction1} is a map from $\mathcal{H}_{\text{past}} \to \mathcal{H}_{\text{future}}$.

\subsubsection{Operator version of soft theorems}

We are now in a position to derive the three-dimensional version of the soft and subleading soft theorems. With the previous notations at hand, again consider the Ward identity $\partial_u \widehat{P} = - \widehat{T}_{ss}^{(-1)}$.  Integrating both sides over $-\infty < s < \infty$, we find
\begin{align}
\label{E:opsoft0}
\widehat{P}(s = \infty, \theta) - \widehat{P}(s = -\infty, \theta) = - \int_{0}^\infty \!\! ds\,\widehat{T}_{ss}^{(-1)}(s,\theta) - \int_{-\infty}^0 \!\! ds\,\widehat{T}_{ss}^{(-1)}(s,\theta)\,,
\end{align}
where the left and right-hand sides are elements of $\mathcal{A}_{\text{tot}} \simeq \mathcal{A}_{\text{future}} \otimes \mathcal{A}_{\text{past}}$.  To make the tensor product between past and future more explicit, we write $\widehat{P}(s = \infty, \theta) = \widehat{P}_f \otimes \mathds{1}$ and $\widehat{P}(s = -\infty, \theta) = \mathds{1} \otimes \widehat{P}_p$, and in a slightly abuse of notation denote $\widehat{T}_{ss}^{(-1)}(s,\theta)$ for $0 \leq s < \infty$ by $\widehat{T}_{ss}^{(-1)}(s,\theta) \otimes \mathds{1}$, and for $-\infty < s \leq 0$ by $\mathds{1} \otimes \widehat{T}_{ss}^{(-1)}(s,\theta)$.  Then~\eqref{E:opsoft0} becomes
\begin{align}
\label{E:opsoftnew}
\widehat{P}_f \otimes \mathds{1} - \mathds{1} \otimes \widehat{P}_p = - \int_{0}^\infty \!\! ds\,\widehat{T}_{ss}^{(-1)}(s,\theta) \otimes \mathds{1} - \int_{-\infty}^0 \!\! ds\,\mathds{1} \otimes \widehat{T}_{ss}^{(-1)}(s,\theta)\,,
\end{align}
and applying the $\widehat{K}$ operator of~\eqref{E:Kop1} we find
\begin{align}
\label{E:boxedsoft1}
\boxed{\widehat{P}_f \cdot \widehat{S} - \widehat{S} \cdot \widehat{P}_p = - \left(\int_{0}^\infty \!\! ds\,\widehat{T}_{ss}^{(-1)}(s,\theta) \cdot \widehat{S} + \int_{-\infty}^0 \!\! ds\,\widehat{S} \cdot \widehat{T}_{ss}^{(-1)}(s,\theta)\right)}
\end{align}
This is the operator form of the soft theorem corresponding to the Ward identity for $P$.  Entirely analogously, we also have
\begin{align}
\label{E:boxedsoft2}
\boxed{\widehat{J}_f \cdot \widehat{S} - \widehat{S} \cdot \widehat{J}_p =  \int_{0}^\infty \!\! ds\,\left(\partial_\theta \!\left(\widehat{P} - \frac{1}{2}\, \widehat{T}_{\theta \theta}^{(0)}\right) - \widehat{T}_{s \theta}^{(-1)}\right) \cdot \widehat{S} + \int_{-\infty}^0 \!\! ds\,\widehat{S} \cdot \left(\partial_\theta \!\left(\widehat{P} - \frac{1}{2}\, \widehat{T}_{\theta \theta}^{(0)}\right) - \widehat{T}_{s \theta}^{(-1)}\right)}
\end{align}
which is the operator form of the soft theorem for the Ward identity for $J$.

Previously we explained how the physical in- and out- states of gravity coupled to massless matter obey constraints, namely the Ward identities~\eqref{eq:boxedEOM} promoted to operator equations that are local on null infinity. The soft theorems are nothing more than the zero-frequency limits of these constraints, together with the antipodal matching condition at spatial infinity.

We can also make contact with the standard derivation of soft theorems in higher dimensions, where one expresses the commutator of the $S$-matrix with the soft charges in terms of matrix elements of hard operators. In this setting the soft charges are $Q_T$ and $Q_Y$, defined in~\eqref{eq:ScalarFieldCharges}, with evolution described by the flux equations~\eqref{eq:QTconservation} and~\eqref{eq:QYconservation}. The hard contributions are simply the matrix elements of the right-hand-side of those flux equations, integrated over all time. For example, the supertranslation soft theorem becomes
\begin{equation}
    \widehat{Q}_{T,f}\cdot \widehat{S} - \widehat{S}\cdot \widehat{Q}_{T,i} = - \left( \int_0^{\infty} \! ds \int_0^{2\pi} \! d\theta \,T(\theta) \widehat{T}_{ss}^{(-1)}(s,\theta)\cdot \widehat{S} + \int_{-\infty}^0 \! ds\int_0^{2\pi}\! d\theta \,T(\theta) \widehat{S}\cdot \widehat{T}_{ss}^{(-1)}(s,\theta) \right)\,,
\end{equation}
with a similar result for the commutator of $\widehat{Q}_Y$ with $\widehat{S}$.

While we have phrased our soft theorems in terms of operator identities, we can equivalently obtain soft theorems for $S$-matrix elements by sandwiching the above operator identities between states in $\mathcal{H}_{\text{future}}$ and $\mathcal{H}_{\text{past}}$.

\subsection{Memory effects}
\label{sec:grav-memory}

Here we compute gravitational memory effects in the presence of ingoing/outgoing massless matter.

\subsubsection{Geodesic deviation equations near asymptotia}

Here we use the same notation for the metric as in Subsection~\ref{subsec:Ward1}.  First, we need to compute the geodesic deviation equation
\begin{equation}
\label{E:deviation1}
(v^\alpha \nabla_\alpha)^2 \xi^\mu = R_{\alpha \beta \gamma}^\mu v^\alpha v^\beta \xi^\gamma\,,
\end{equation}
at large $r$, where $v^\mu$ is the three-velocity of a timelike geodesic and $\xi^\mu$ is the deviation vector.  We proceed by expanding $v^\mu$ at large $r$ using our asymptotic expansion~\eqref{eq:ScalarBCs} and~\eqref{E:asym2}, and then solve~\eqref{E:deviation1} for the asymptotic behavior of the deviation $\xi^\mu$. The velocity vector $v^\mu$ satisfies the normalization condition $v_\mu v^\mu = -1$, and since $v^\mu$ is the velocity of a timelike geodesic the geodesic equation enforces $v^\nu \nabla_\nu v^\mu  = 0$.  The two previous equations can be solved at large $r$ by
\begin{align}
\begin{split}
v^u &= 1 - \frac{2(L_{-1}(u,\theta) - L_{-1}(u_0,\theta))}{r} +O(r^{-2})\,,
\\
v^r &= \frac{1}{2}(1 + M_1(u,\theta)) 
\\
& \qquad + \frac{\frac{1}{2}\,M_0(u,\theta) - M_1(u,\theta)\left(L_{-1}(u,\theta) - L_{-1}(u_0, \theta)\right) - L_{-1}(u_0, \theta)}{r} +O(r^{-2})\,,
\\
v^\theta &= \frac{\frac{1}{2}\int_{u_0}^u dw \, \partial_\theta M_1(w,\theta) - (N_{-2}(u,\theta) - N_{-2}(u_0,\theta))}{r^2}+O(r^{-3})\,.
\end{split}
\end{align}

Next we expand $\xi^\mu$ at large $r$ as
\begin{align}
\begin{split}
\xi^u(u,r,\theta) &= \xi_0^u(u,\theta) + \frac{\xi_{-1}^u(u,\theta)}{r} + \frac{\xi_{-2}^u(u,\theta)}{r^2} + O(r^{-3})\,, \\
\xi^r(u,r,\theta) &= \xi_0^r(u,\theta) + \frac{\xi_{-1}^r(u,\theta)}{r} + \frac{\xi_{-2}^r(u,\theta)}{r^2} +O(r^{-3})\,, \\
\xi^\theta(u,r,\theta) &= \xi_0^\theta(u,\theta) + \frac{\xi_{-1}^\theta(u,\theta)}{r} + \frac{\xi_{-2}^\theta(u,\theta)}{r^2} + O(r^{-3})\,.
\end{split}
\end{align}
At leading order in large $r$, the geodesic deviation equation~\eqref{E:deviation1} gives $\partial_u^2 \xi_0^u = 0$ and $\partial_u^2 \xi_0^\theta = 0$, and so we can write
\begin{align}
\begin{split}
\xi_0^u(u,\theta) &= u\,\dot{\xi}_0^u(\theta) + \xi_0^u(\theta) \\
\xi_0^\theta(u,\theta) &= u\,\dot{\xi}_0^\theta(\theta) + \xi_0^\theta(\theta)\,.
\end{split}
\end{align}
Then there is a non-trivial equation for $\xi_0^r(u,\theta)$ whose solution is
\begin{align}
\label{E:fullmemory1}
\begin{split}
&\xi_0^r(u_f ,\theta) - \xi_0^r(u_i ,\theta) = \int_{u_i}^{u_f} \!du' \int_{u_0}^{u'} \! du \Big(\dot{\xi}_0^\theta(\theta) \,\partial_\theta M_1(u,\theta) + \dot{\xi}_0^u(\theta) \,\partial_u M_1(u,\theta) \\
& \qquad \qquad \qquad \qquad \qquad \qquad \qquad \qquad \quad + \frac{1}{2}\,\xi_0^\theta(u,\theta) \,\partial_u \partial_\theta M_1(u,\theta) + \frac{1}{2}\,\xi_0^u(u,\theta) \,\partial_u^2 M_1(u,\theta)\Big)
\end{split}
\end{align}
under the assumption that $\partial_u\xi_0^r(u,\theta)\big|_{u = u_0} = 0$.  The above simplifies for various choices of initial conditions.  For instance, if we suppose that $\dot{\xi}_0^u(\theta) = \dot{\xi}_0^\theta(\theta) = 0$ and that $M_1(u_0,\theta) = -1$, then~\eqref{E:fullmemory1} becomes
\begin{align}
\label{E:deviationboxed1}
\xi_0^r(u_f ,\theta) - \xi_0^r(u_i ,\theta) =  \frac{1}{2}\int_{u_i}^{u_f} du \left(\xi_0^\theta(\theta) \,\partial_\theta M_1(u,\theta) + \xi_0^u(\theta) \,\partial_u M_1(u,\theta)\right).
\end{align}
Assuming our spacetime is sourced by a stress tensor, writing $M_1 = 16 \pi G\,P$ and $N_{-2} = 8 \pi G\,J$ as before, we can use the Ward identities~\eqref{eq:boxedEOM} to simplify~\eqref{E:deviationboxed1} as
\begin{align}
\label{E:memorybox1}
\boxed{\Delta \xi_0^r(\theta) = 8 \pi G\,\xi_0^\theta(\theta) \left(\Delta J(\theta) + \int_{u_i}^{u_f}du \left(\frac{1}{2}\,\partial_\theta T_{\theta \theta}^{(0)} + T_{u\theta}^{(-1)}\right)\right) - 8 \pi G\,\xi_0^u(\theta)\int_{u_i}^{u_f} du\,T_{uu}^{(-1)}}
\end{align}
where $\Delta \xi_0^r(\theta)$ is a shorthand for $\xi_0^r(u_f ,\theta) - \xi_0^r(u_i ,\theta)$ and $\Delta J(\theta)$ is a shorthand for $J(u_f,\theta) - J(u_i,\theta)$.  Hence we see a ``rotational displacement'' memory effect and a ``time displacement'' memory effect visible at leading order in $1/r$.  These memory effects are both sourced in part by integrated null energy depositions.  Unlike in four spacetime dimensions these memory effects describe a permanent radial displacement rather than angular displacements.  Part of the difference from the four-dimensional case is that the boundary conditions~\eqref{E:asymptoticallyflat1} of 3d gravity allow for larger falloffs.

\subsubsection{Shockwave sources}
\label{Subsec:shock1}

Here we show that the form of the asymptotically flat space metric~\eqref{eq:ScalarBCs} (see also~\eqref{E:asym2}) holds in the presence of a shockwave. To this end, consider the shockwave metric
\begin{equation}
\label{E:shockwavemetric1}
ds^2 = - (1 -  16 \pi G \,\widetilde{P}(\theta)\,\Theta(u - u_s)) \,du^2 - 2 \,du \, dr + 16 \pi G\,\widetilde{J}(u,\theta) \,\Theta(u-u_s)\, du \, d\theta + r^2  d\theta^2\,,
\end{equation}
with the relation
\begin{equation}
\partial_u \widetilde{J}(u,\theta) = \partial_\theta\widetilde{P}(\theta)\,.
\end{equation}
We note that $\widetilde{P}, \widetilde{J}$ are related to $P,J$ by $P(u,\theta) = \widetilde{P}(\theta) \,\Theta(u - u_s)$ and $J(u,\theta) = \widetilde{J}(u,\theta)\,\Theta(u - u_s)$.
Notice that the Einstein field equations give us
\begin{equation}
\label{E:shockstress1}
T_{uu} = \delta(u-u_s) \left(- \frac{\widetilde{P}(\theta)}{r} +\frac{\partial_\theta \widetilde{J}(u,\theta)}{r^2} - \frac{4 \pi G\,\widetilde{J}(u,\theta)^2}{r^3}\right) \,, \qquad T_{u \theta} = - \delta(u-u_s)\, \frac{\widetilde{J}(u,\theta)}{r}\,,
\end{equation}
with all other components zero.  Evidently, the stress tensor is supported at $u = u_s$, and so indeed~\eqref{E:shockwavemetric1} manifests the geometry of flat space in the presence of a null shockwave. It is readily checked that
\begin{align}
\begin{split}
\beta(u,r,\theta) &= 0 \,,
\\
V(u,r,\theta) &= -\left(1 - 16\pi G\,\widetilde{P}(\theta) \, \Theta(u-u_s)\right)\,r - \frac{(8 \pi G\,\widetilde{J}(u,\theta))^2\,\Theta(u-u_s)}{r}\,,
\\
U(u,r,\theta) &= -\frac{8\pi G\,\widetilde{J}(u,\theta)\,\Theta(u-u_s)}{r^2}\,,
\end{split}
\end{align}
in accordance with the far-field constraints~\eqref{E:asym2}. These results are consistent with the BMS Ward identities~\eqref{eq:boxedEOM}. In particular, the shockwave  injects supermomentum and super-angular momentum at time $u_s$. There are some interesting subtleties in the above computations pertaining to the presence of the theta and delta functions. See Appendix~\ref{App:comments} for some comments.

Using~\eqref{E:fullmemory1} and taking $u_0 < u_s$ as well as $u_i < u_s < u_f$ and $\dot{\xi}_0^\theta(\theta) = 0$, we find the memory effect
\begin{align}
\Delta \xi_0^r(\theta) = 8 \pi G\,\xi_0^\theta(\theta) \left(\Delta J(\theta) - \widetilde{J}(u_s,\theta)\right) + (u_f - u_s)\,8 \pi G\,\dot{\xi}_0^u(\theta) \,\widetilde{P}(\theta)\,.
\end{align}

\section{Discussion}
\label{sec:discussion}

There has been some expectation, from the point of view of soft theorems and memory effects, that three-dimensional flat space gravity is in some way trivial. Our work has shown that this is not the case. This simple model has boundary gravitons described by a soluble quantum field theory whose role is to implement BMS$_3$ Ward identities and their selection rules. Otherwise, for the quantities we have computed, the physics of the soft gravitons decouple. In totality, three-dimensional gravity is not as barren as was previously anticipated: it furnishes an infinite-dimensional Hilbert space of soft vacua, infinite-dimensional asymptotic symmetries, memory effects, and soft theorems. In this sense three-dimensional gravity shares in common more features with its four-dimensional counterpart than was previously appreciated. 

Boundary degrees of freedom are usually required to implement selection rules in quantum field theory.  In our setting the boundary gravitons are the degrees of freedom implementing $\widehat{\text{BMS}}_3$ selection rules. It is reasonable to expect similar boundary degrees of freedom to implement soft theorems in gauge theories and gravity theories more generally, including in the physically relevant case of four dimensions. We suspect that these needed degrees of freedom can be found by a careful path integral analysis of boundary terms in the effective action and their interplay with boundary conditions required to render a well-posed variational principle. See Refs.~\cite{Kapec:2022hih,Nguyen:2020hot, Nguyen:2021ydb,Barnich:2022bni,He:2024vlp} for promising work in this vein. 

Let us conclude with some comments on the connection of our work to the celestial holography program.  Conformal Carrollian field theories may provide putative bulk duals to flat space quantum gravity akin to the AdS/CFT correspondence, and their correlation functions are related to celestial CFTs via integral transform~\cite{Donnay:2022aba, Bagchi:2022emh, Donnay:2022wvx, Kim:2023qbl, Jain:2023fxc}. In a recent work~\cite{Cotler:2024xhb}, we studied the quantization of Carrollian field theories, including both `pure electric' theories, `pure magnetic' theories, and coupled `electromagnetic' theories.  We argued that two-derivative theories with an electric sector have UV/IR mixing which necessitates a multiplicative renormalization scheme for couplings. In the present paper, the boundary description obtained above is a `pure magnetic' conformal Carrollian theory, devoid of UV/IR mixing. 

We can regard this boundary system as a conformal Carrollian field theory for the stress tensor.  The theory is a description of the hydrodynamics of the stress tensor, capturing its conservation law and current algebra more broadly.  While our theory is an exact description of pure flat space three-dimensional gravity, it should be regarded as an effective description of the stress tensor sector of a putative conformal Carrollian field theory describing the microscopic completion of three-dimensional gravity coupled to matter. This is analogous to the status of the Schwarzian theory for nearly-AdS$_2$ gravity~\cite{Jensen:2016pah, Maldacena:2016upp, Engelsoy:2016xyb}, or the Alekseev-Shatashvili theory for AdS$_3$ gravity~\cite{cotler2019theory}.  From this point of view, we have discovered a universal sector of putative conformal Carrollian field theories for three-dimensional gravity.

\subsection*{Acknowledgments}

It is a pleasure to thank Glenn Barnich, Prateksh Dhivakar, Lorenz Eberhardt, Hernán González, Daniel Grumiller, Blaza Oblak, and Andrew Strominger for enlightening conversations. JC is supported by the Simons Collaboration on Celestial Holography. KJ is supported in part by an NSERC Discovery Grant. JS is supported by a postdoctoral research fellowship of the
F.R.S.-FNRS (Belgium).

\appendix 

\section{Comments on products of $\delta$ distributions and $\theta$ functions}
\label{App:comments}

In shockwave calculations for Einstein gravity, we often encounter distributional terms including Heaviside Theta functions, Dirac delta functions, and composites built from the two. This leads to potential ambiguities that must be addressed when solving the Einstein equations. 

In the shockwave computation presented in the main text in Subsection~\ref{Subsec:shock1} we encountered the product $\Theta(x)\delta(x)$. We begin by carefully defining this distribution. We should regularize $\Theta(x)$ through a function $\theta_\varepsilon(x)$ which becomes the $\Theta(x)$ function as $\varepsilon \to 0$. We choose $\theta_{\varepsilon}$ such that it monotonically increases and its derivative has compact support $(-\varepsilon,\varepsilon)$. In the shockwave calculations, the $\delta$ distribution always arises as a derivative of the $\theta$ function, so we define the regularization $\delta_\varepsilon(x) := \theta_\varepsilon'(x)$. Now let $f(x)$ be a $C^{\infty}$ test function in a neighborhood $(-a,b)$ of the origin. Then
\begin{align}
\begin{split}
    \int_{-a}^b dx\, \delta_\varepsilon(x) \, \theta_\varepsilon(x) \, f(x) &=  \int_{-\varepsilon}^{\varepsilon} dx \,f(x)\, \delta_\varepsilon(x) \, \theta_\varepsilon(x) =  \int_{-\varepsilon}^{\varepsilon} dx \,f(x) \frac{d}{dx}\left( \frac{1}{2}\,\theta_\varepsilon(x)^2 \right)
    \\
    &= \left( f(x) \,\frac{1}{2}\,\theta_\varepsilon(x)^2\right)\Big|_{-\varepsilon}^{\varepsilon} - \int_{-\varepsilon}^{\varepsilon}dx\,f'(x)\frac{1}{2}\theta_{\varepsilon}(x)^2 \,.
\end{split}
\end{align}
The integral in the second line vanishes as $\varepsilon\to 0$ while the boundary term becomes $\frac{1}{2}f(0)$. 

The above means that in this context it is appropriate to identify $\delta(x) \, \Theta(x)$ with $\frac{1}{2}\,\delta(x)$.  Note, however, that the $\frac{1}{2}$ does \textit{not} arise because of a convention for $\Theta(0)$, say assigning it the value $ \frac{1}{2}$. Rather, it comes from taking the antiderivative of $\delta_\varepsilon(x)\,\theta_\varepsilon(x)$ which is $\frac{1}{2}\,\theta_\varepsilon(x)^2$. Similar logic shows that when integrated against smooth test functions the ambiguous expression 
$\delta(x) \Theta(x)^n$ can be understood as $\frac{1}{n+1}\delta(x)$. Relatedly, this tells us that we should not always identify $\Theta(x)^n$ as $\Theta(x)$.

A few more comments are in order. The logic above hinges on the identification $\delta_\varepsilon(x) := \theta_\varepsilon'(x)$.  This is justified in a shockwave calculation for Einstein gravity: the $\delta$ distributions arise as derivatives of a single progenitor $\theta$ function. More broadly the identification $\delta_\varepsilon(x) := \theta_\varepsilon'(x)$ seems sensible usually, but not always. As an example, let $a \in (0,1)$, $g(x) = \frac{1}{2} + \frac{1}{2}\,\tanh(x)$, and $h(x) = \frac{1}{2} + \frac{1}{2}\, \text{erf}(x)$, and define
\begin{align}
\begin{split}
\label{E:alternateThetaDelta}
\theta_\varepsilon(x) &:= g\!\left(\frac{x}{\varepsilon} + g^{-1}(a)\right) \,,\\
\delta_\varepsilon(x) &:= \frac{1}{\varepsilon^2}\,h'\!\left(\frac{x}{\varepsilon^2}\right)\,.
\end{split}
\end{align}
Here $\delta_\varepsilon(x)$ is not the derivative of $\theta_\varepsilon(x)$. It can be checked that with these definitions and for a $C^{\infty}$ function $f(x)$ on the interval $(-\alpha,\beta)$ including the origin,
\begin{equation}
\lim_{\varepsilon \to 0} \int_{-\alpha}^{\beta} dx\,f(x) \delta_\varepsilon(x)\,\theta_\varepsilon(x) = a f(0)\,,
\end{equation}
and hence we can interpret $\delta(x)\Theta(x)$ to be any value in $(0,1)$ times $\delta(x)$ by tuning $a$. This particular limiting form of the $\Theta$ functions obeys $\theta_\varepsilon(0) = a$ for all $\varepsilon$, and relatedly we have
\begin{equation}
\lim_{\varepsilon \to 0} \int_{-\alpha}^{\beta} dx\,f(x) \delta_\varepsilon(x)\,\theta_\varepsilon(x)^n = a^n f(0)\,,
\end{equation}
and so for our regularization we can replace $\delta(x) \, \Theta(x)^n$ by $\delta(x) \, \Theta(0)^n = a^n \, \delta(x)$. We emphasize that these results depend on the particular definitions of $\theta_\varepsilon(x)$ and $\delta_\varepsilon(x)$ in~\eqref{E:alternateThetaDelta}.

Now we explain some general rules that should apply to most physics problems with ambiguous distributions. \\ \\
\textbf{Rules for when it is sensible to choose a regularization such that $\delta_\varepsilon(x) = \theta_\varepsilon'(x)$.}
\begin{enumerate}
\item Retain powers of $\Theta(x)^n$ without simplification.
\item Set $\Theta'(x) = \delta(x)$ at intermediate steps.
\item At the very end of the calculation, one can set $\Theta(x)^n = \Theta(x)$ for $n \in \mathbb{Z}_{\geq 1}$\,, and $\delta(x) \, \Theta(x)^n = \frac{1}{n+1}\, \delta(x)$ for $n \in \mathbb{Z}_{\geq 0}$\,.
\end{enumerate}
\noindent We conclude by noting that there is an algebraic formalization of the above ideas, called the Colombeau algebra~\cite{colombeau2011elementary, gratus2013colombeau}.  In practice, the above rules suffice for most problems in physics.

\section{Comment on Chern-Simons-matter theory and geometric actions}
\label{App:absence}

Chern-Simons theory on a topological solid cylinder $M$ with suitable boundary conditions can be rewritten in terms of chiral Wess-Zumino-Witten theory on the boundary $\partial M$. The chiral WZW action is a geometric action describing motion on a coadjoint orbit of the underlying gauge group. These statements can be demonstrated in a very similar way to the derivation of the boundary graviton action of flat space gravity described in Section~\ref{sec:bdyQFT}.

Now consider Chern-Simons-matter theory. For simplicity, we can take a free massless charged scalar coupled to $U(1)_k$ Chern-Simons theory on a solid cylinder. Picking a time foliation we can write the action in Hamiltonian form as
\begin{equation}
    \int dt d^2x \left( \frac{k}{4\pi} \varepsilon^{ij} A_i \dot{A}_j + \pi \dot{\phi}+\bar{\pi}\dot{\bar{\phi}}- |\pi|^2 - |(\partial_i - i A_i) \phi|^2 +A_0\left( \frac{k}{2\pi}\varepsilon^{ij}F_{ij}- \rho \right)\right) +(\text{bdy})\,,
\end{equation}
where the charge density is $\rho= i (\bar{\pi}\bar{\phi}-\pi \phi)$. Classically, $A_0$ acts as a Lagrange multiplier fixing the spatial field strength in terms of the charge density through the Gauss' Law constraint. Quantum mechanically this remains true in Coulomb gauge, since the gauge-fixing auxiliary field and Faddeev-Popov ghosts do not couple to $A_0$. Given a matter configuration, the Gauss' Law constraint can in principle be solved to give $A_i$ up to a large gauge transformation, which remains a dynamical degree of freedom after enforcing the constraint. For example, in three-dimensional flat space in a mixed time-momentum representation, the Gauss' Law constraint implies
\begin{equation}
    \widetilde{A}_i (t,\vec{k}) = i \frac{2\pi}{k}\frac{ \varepsilon_{ij}k^j}{k^2}\tilde{\rho}(t,\vec{k}) + i k_i \widetilde{\Lambda}(t,\vec{k})\,,
\end{equation}
where $\widetilde{\Lambda}$ denotes the large gauge transformation. In the presence of a lump of matter of compact support $A_i$ is pure gauge $\Lambda$, at the boundary with a holonomy determined by the total bulk charge. In the absence of matter, or if the matter is a timelike Wilson line, then one can reduce the bulk action to a boundary to one for the edge mode $\Lambda|_{\partial M}$. If the matter is a network of Wilson lines then the configuration of $\Lambda$ is usually specified by bulk moduli as well as the edge modes. For dynamical matter however, we cannot integrate out the bulk modes of $A_i$ and thereby obtain a local boundary action for the edge mode. Perhaps the main obstruction is that the holonomy for $\Lambda|_{\partial M}$, fixed in terms of the total off-shell bulk charge, is a fluctuating and nonlocal variable. On-shell the total charge only changes when injected or removed from the boundary, but in the path integral it is a fluctuating variable. For this reason, it is only sensible to obtain a geometric action for the edge modes when the matter can be treated classically.

\bibliographystyle{utphys}
\bibliography{bibl}

\end{document}